\documentclass[preprint,nofootinbib]{revtex4-1}

\usepackage{graphicx}  % needed for figures
\usepackage{dcolumn}   % needed for some tables
\usepackage{bm}        % for math
\usepackage{amssymb}
\usepackage{diagbox}
\usepackage{slashbox}
\usepackage{slashbox}
\usepackage{hyperref}
\usepackage{natbib}
\usepackage[table]{xcolor} % Load xcolor with table support
\usepackage{pifont}
\usepackage{multirow}
\usepackage{amsmath}
\usepackage{cases}
\usepackage{placeins}
\usepackage{textcomp}
\UseRawInputEncoding

\usepackage{blindtext}
\usepackage{url}
\usepackage{graphicx}
\usepackage{url}
\usepackage{tikz}
\usepackage[bf]{caption}
\usepackage{amsmath}
\usepackage{cases} 
\usepackage{multirow}

\usepackage{epsfig}
\usepackage{makecell}
\usepackage{amsmath}
\usepackage{amsfonts}
\usepackage{amssymb}
\usepackage{mathtools}
\usepackage{url}
\usepackage{subfigure}
\usepackage{hhline}
\usepackage{color}
\usepackage{array,multirow}
\usepackage{bm}
\usepackage{makecell}
\usepackage{epsfig}
\usepackage{amsmath}
\usepackage{amsfonts}
\usepackage{amssymb}
\usepackage{url}
\usepackage{hyperref}
\usepackage{subfigure}
\usepackage{hhline}
\usepackage{color}
\usepackage{bm}
\usepackage{cancel}
\usepackage{mathtools}

\newcommand{\beqa}{\begin{eqnarray}}
\newcommand{\eeqa}{\end{eqnarray}}
\newcommand{\beq}{\begin{equation}}
\newcommand{\eeq}{\end{equation}}

\newcommand{\nn}{\nonumber}
\newcommand{\bmt}{\begin{pmatrix}}
\newcommand{\emt}{\end{pmatrix}}
\usepackage[toc,page]{appendix}
\usepackage{comment}
\newcommand{\be}{\begin{equation}}
\newcommand{\ee}{\end{equation}}
\newcommand{\bea}{\begin{eqnarray}}
\newcommand{\eea}{\end{eqnarray}}

\hypersetup{
    colorlinks=true,
    citecolor=red,
    linkcolor=blue,
    filecolor=green,      
    urlcolor=magenta,
}

\begin{document}
\title{\bf Analysis of  $b \to c \ell \nu $ baryonic decay modes in  SMEFT approach}
%%%%%%%%%%%%%%%% Exploring $b \to c \ell \nu $ baryonic decay mode in SMEFT approach
\author{Dhiren Panda}
\email{pandadhiren530@gmail.com}

\author{Manas Kumar Mohapatra}
\email{manasmohapatra12@gmail.com}

\author{Rukmani Mohanta}
\email{rmsp@uohyd.ac.in}
\affiliation{School of Physics,  University of Hyderabad, Hyderabad-500046,  India}

%%%%%%%%%%%%%%%%%%%%%%%%%%%%%%%%%%%%%%%%%%%%%%%%%%%%%%%%%%%%
\begin{abstract}
%%%%%%%%%%%%%%%%%%%%%%%%%%%%%%%%%%%%%%%%%%%%%%
% The flavor-changing neutral current (FCNC) processes in hadronic $B$ decays, specifically those mediated by the $b \to s \ell \ell$ transition, offer a powerful avenue to explore physics beyond the Standard Model (SM). In this work, we investigate the new physics impact within the framework of the Standard Model Effective Field Theory (SMEFT). By analyzing the charged current $b \to c \ell \nu$ process in conjunction with the neutral current $b \to s \ell \ell$ decays, we study complementary insights into possible new physics effects.
% We focus on the feasibility of interpreting the processes mediated by $b \to c \tau \nu$ transitions, in particular, the semileptonic $b$-baryonic decay modes $\Sigma_b \to \Sigma_c^{(*)} \tau^-\bar{\nu}\tau$ and $\Xi_b \to \Xi_c \tau^-\bar{\nu}\tau$ in the context of new physics scenarios. In the SMEFT approach, we investigate the sensitivity of new physics operators on various observables such as branching ratio, forward-backword asymmetry, the lepton non-universal observable, the longitudinal polarization fractions of the $b$-baryonic decay channels.  
%%%%%%%%%
The flavor-changing neutral current decays of heavy bottom quark, alongside the flavor-changing charged current processes mediated by $b \to (c, u)$ in semileptonic $B$ decays are emerged as powerful tools for exploring physics beyond the Standard Model. In this work, we focus on the feasibility of interpreting the processes mediated by $b \to c \tau \nu$ transitions, in particular, the semileptonic $b$-baryonic decay modes $\Sigma_b \to \Sigma_c^{(*)} \tau^-\bar{\nu}_\tau$ and $\Xi_b \to \Xi_c \tau^-\bar{\nu}_\tau$ in the context of SMEFT approach. We perform a detailed analysis of the sensitivity of new physics operators on various observables such as branching ratio, forward-backward asymmetry parameter,  lepton non-universal observable and the longitudinal polarization fraction of the $b$-baryonic decay channels.  
%%%%%%%
\end{abstract}
%%%%%%%%%%%%%%%%%%%%%%%%%%%%%%%%%
\maketitle
%%%%%%%%%%%%%%%%%%%%%%%%%%%%%%%%%%%%%%%%%%%%%%%%%%%%%%%%%%%%%%%%%%%%%%%%%%%%%%%%
\section{Introduction}
 In recent years, $b$-hadron decays have garnered significant attention as a promising avenue for exploring physics beyond the Standard Model (BSM). Both neutral and charged current $b$-decays offer a clean and controlled environment to investigate the sensitivity of new physics. Notably, the violation of Lepton Flavor Universality (LFU) - the principle that leptons couple to gauge bosons in a flavor-independent manner, has been observed in several observables in the semileptonic $B$ decays. The observables associated with the flavor changing charge current (FCCC) processes such as $R_{D^{(*)}}$, defined as
 %%%%%%%%%%%5
\bea
R_{D^{(*)}}=\frac{\mathcal{B}(B \to D^{(*)} \tau \nu)}{\mathcal{B} (B \to D^{(*)} \ell \nu)},
\eea
with ($\ell = \mu, e$), are potentially sensitive to the lepton flavor universality violation, making them ideal tools for testing new physics effects.
%%%%%%%%%%%%5
Several measurements on $R_{D^{(*)}}$ at BaBar \cite{BaBar:2012obs, BaBar:2013mob}, Belle \cite{Belle:2015qfa, Belle:2016ure, Abdesselam:2016xqt, Belle:2019rba}, and LHCb \cite{LHCb:2015gmp, LHCb:2017smo} collaborations along with HFLAV Group \cite{HFLAV:2022esi} show deviation of approximately $3.3\sigma$  from the Standard Model (SM) prediction \cite{Fajfer:2012vx, Kamenik:2008tj, Aoki:2016frl}.
Other observables, such as $P_{\tau}(D^{})$ and $F_{L}(D^{})$, which involve the ratio of $B$ meson decays for polarized and unpolarized final states, exhibit deviations from SM predictions at the $(1.5-2)\sigma$ level \cite{Belle:2016dyj, Asadi:2018sym, Alok:2016qyh}. Numerous global analyses have explored the tension in $R_{D^{(*)}}$ by incorporating new physics (NP) contributions. Specifically, various one and two-dimensional NP   scenarios have been proposed as plausible solutions, with particular emphasis on new physics affecting only the $\tau$ channel \cite{Ray:2023xjn}. Similarly, the ratio of the branching fractions $R_{J/\psi}=\mathcal{B}(B_c \to J/\psi \tau \nu)/\mathcal{B} (B \to J/\psi \ell \nu)$ measured by LHCb collaboration \cite{LHCb:2017vlu} also indicates about $2\sigma$ discrepancy above the SM value \cite{Dutta:2017xmj}.
%%%%%%%%%%%%%%%%%%%

Motivated by the insights from the above decay processes, we investigate semileptonic $b$-baryon decays, specifically focusing on the exclusive $\Sigma_b \to \Sigma_c^{(*)} \tau^-\bar{\nu}_\tau$ and $\Xi_b \to \Xi_c \tau^-\bar{\nu}_\tau$ processes. These decay modes could offer valuable information for probing new physics effects in the context of $b \to c \ell \nu$ quark-level transitions. Investigation of these decays offers crucial insights into the weak interaction properties and underlying dynamics of heavy baryons, as well as their sensitivity to NP. Furthermore, semileptonic decays of heavy $b$-baryons provide a valuable complementary perspective to meson decay modes, enriching the search for new physics by exploring different aspects of particle interactions. On the experimental front, significant progress has been made in studying heavy baryons containing a $b$ quark with substantial data accumulation from experiments like the Tevatron and LHC. The $\Xi_b^{0,-}$ baryons have been observed by the CDF, D0 \cite{CDF:2007cgg, D0:2007gjs}, and LHCb \cite{LHCb:2014chk, LHCb:2014wqn, LHCb:2013jih, LHCb:2014jst, LHCb:2015une, LHCb:2016hha, LHCb:2017fwd} collaborations. 
Additionally, the CDF and LHCb collaborations \cite{CDF:2011ac, LHCb:2018haf} have reported clear signals for the four strongly decaying baryon states: $\Sigma_b^{+}$, $\Sigma_b^{-}$, $\Sigma_b^{*+}$, and $\Sigma_b^{*-}$.   While the $\Sigma_b$ baryon primarily decays through strong interactions, measuring its weak branching fraction can be challenging. Yet, investigating semileptonic decay modes may yield significant insights, as they exhibit a higher sensitivity to departures from SM predictions, potentially pointing to the new physics \cite{Ke:2019smy}.
%%%%%%%%%%%%%%%%%%%

In this work, we adopt the Standard Model Effective Field Theory (SMEFT) framework, assuming that the scale of NP exceeds the electroweak scale and that no additional light particles exist below this scale. Given the lack of direct evidence for new particles, a model-independent effective field theory approach is ideal. Within this framework, the low energy effective field theory (LEFT) can describe the $b \to c \ell \nu$ transition below the electroweak scale, respecting SM gauge symmetry $SU(3)_C \times SU(2)_L \times U(1)_Y$.
%%%%%%%%%%%%%%%%%%%%%%
We assume the new physics scale $\Lambda$ to be 1 TeV, which is much higher than the electroweak (EW) scale. In this context, the physics between the EW scale and $\Lambda$ can be effectively described by the Standard Model Effective Field Theory (SMEFT), which includes all SM particles. 
% The SM can be considered an effective theory of a more fundamental underlying theory, with new physics effects captured by higher-dimensional operators in SMEFT.
%%%%%%%%%%%%%%%%
To connect low-energy observables with new physics, it is essential to match the operators in the LEFT to those in SMEFT and then run down to $m_b$ scale. The SMEFT operators that generate the $b \to c \ell \nu_{\ell}$ transition also produce LEFT operators relevant for the $b \to s \ell^{+} \ell^{-}$ and $b \to s \nu \nu$ processes.
%%%%%%%%%%
This connection means that the constraints from $b \to c \ell \nu$ processes are inherently tied to bounds from $b \to s \ell \ell$ and $b \to s \nu \nu$ processes. Therefore, the limits derived from $b \to s \ell^+ \ell^-$ and $b \to s \nu \nu$ transitions serve as complementary constraints on the decay modes being investigated. Various theoretical analyses have been conducted to explore the baryonic/mesonic $b \to c \ell \nu$ quark decay processes in detail \cite{Blanke:2018yud,Sahoo:2016pet,Fedele:2022iib,Calibbi:2015kma,Huang:2018nnq, Blanke:2019qrx,Ebert:2006rp, Singleton:1990ye,Crivellin:2018yvo,Crivellin:2022qcj, Ivanov:1996fj,Das:2019cpt, Das:2021lws, Ivanov:1998ya, Ke:2012wa, Wang:2017mqp, Rajeev:2019ktp, Capdevila:2017iqn,Das:2023gfz,Sahoo:2019hbu,Ray:2019gkv,Colangelo:2024mxe,Colangelo:2020vhu} providing valuable
insights into their underlying dynamics. We mainly focus on the various decay observables such as branching ratio, forward-backward asymmetry, covexity parameter\textcolor{red}{,} and lepton polarisation asymmetry of the exclusive $\Sigma_b \to \Sigma_c^{(*)} \tau^-\bar{\nu}_\tau$ and $\Xi_b \to \Xi_c \tau^-\bar{\nu}_\tau$ transitions. We also scrutinize the nature of the lepton non-universality of the above decay channels.
%%%%%%%

%%%%%%%%%%%%%

This paper is organized as follows. In section \ref{subsec:theory}, we will first introduce the LEFT to describe the $b \to c \tau \nu $ process and eventually match with the SMEFT operators.  In section \ref{sec:phen}, we discuss various observables associated with the exclusive $\Sigma_b \to \Sigma_c^{(*)} \tau^-\bar{\nu}_\tau$ and $\Xi_b \to \Xi_c \tau^-\bar{\nu}_\tau$ processes. Section \ref{new physics} focuses on the new physics analysis, including the constraint on the NP couplings within both one-dimensional and two-dimensional scenarios. We then examine the effect of these newly fitted couplings on the $b \to c \tau \bar{\nu}$ baryonic decay observables in section \ref{result and discussion}. Finally, we conclude our analysis in section \ref{sec:conclu}.
%%%%%%%%%%%%%%%%%%%%%%%%%%%%%%%%%%%%%%%%%%%
%\section{The working frame of effective field theory}
\section{Theoretical Framework}
\label{subsec:theory}
 This section outlines the framework for the $b \to c \ell \nu$ process, along with other relevant transitions such as $b \to s \ell \ell$ and $b \to s \nu \nu$ within the LEFT. 
 It also contains the details of the operators relevant to those in the SMEFT at $\mu(m_b)$ scale. In this work, we discuss the effects of the NP couplings, assuming that all Wilson coefficients (WCs) are real.

\subsection{ Low Energy Effective Field Theory}
\subsubsection{$b \to c \ell \nu$}
The most general low-energy effective Hamiltonian relevant for $b \to c \ell \nu $ transition, considering only the left-handed neutrinos, is given as~\cite{Sakaki:2013bfa}
\begin{equation}\label{eq:234}
     \begin{split}
         \mathcal H_{eff} =\mathcal{H}_{eff}^{SM}+\frac{4G_F}{\sqrt{2}}V_{cb}\sum_{i,l}C_i^{(l)} \mathcal{O}_i^{(l)}+h.c.
\end{split}
 \end{equation}
 Here $\mathcal{O}^{(\ell)}_i$ ($i=V_L, V_R, S_L, S_R, T$) are local effective operators, $C_i^{(\ell)}$ are Wilson coefficients encoding contributions of short distance new physics, $V_{cb}$ is the Cabbibo-Kobayasi-Maskawa (CKM) matrix element,  $G_{F}$ is the Fermi constant, and $\ell$ stands for the lepton flavor indices. The relevant set of local operators with mass dimension-6 are presented as
\bea
&& O_{V_L}^{(\ell)} =
(\bar{c}_L \gamma^{\mu} b_L)(\bar{\ell}_L \gamma_\mu \nu_{\ell L}),
\hspace{1cm} O_{V_R}^{(\ell)} =
(\bar{c}_R \gamma^{\mu} b_R)(\bar{\ell}_L \gamma_\mu \nu_{\ell L}),\nn\\
&& O_{S_L}^{(\ell)} =
(\bar{c}_R b_L)(\bar{\ell}_R \nu_{\ell L}),
\hspace{1.7cm}
O_{S_R}^{(\ell)} =
(\bar{c}_L b_R)(\bar{\ell}_R \nu_{\ell L}),\nn\\
&&
O_{T}^{(\ell)} =
(\bar{c}_R \sigma^{\mu\nu} b_L)(\bar{\ell}_R \sigma_{\mu\nu} \nu_{\ell L}),
\label{eq:effopr}
\eea 
where the $O_{V_{L,R}}^{(\ell)}$, $O_{S_{L,R}}^{(\ell)}$, and $O_{T}^{(\ell)}$ are respectively known as vector, scalar, and tensor operators.
%%%%%%%%%%%%%%%%%%%%%%%%%%$
\subsubsection{$b \to s \ell \ell$}
The weak effective Hamiltonian describing the $b \to s\ell \ell$ decay  is given as \cite{Bobeth:1999mk, Bobeth:2001jm}
 \begin{equation}\label{eq:234}
     \begin{split}
         \mathcal H_{eff} =-\frac{4G_{F}}{\sqrt{2}}V_{tb}V^{*}_{ts}\frac{\alpha_{em}}{4\pi}\sum_{i=9, 10,S,P}\big(C_{i}\mathcal{O}_{i}+C_{i}\mathcal{O'}_{i}\big),  
\end{split}
 \end{equation}
where $V_{tb}V_{ts}^*$ is the product of the CKM matrix elements, and $\alpha_{em}$ is the fine structure constant.
The relevant operators for this process are expressed as follows,
 \begin{eqnarray}\label{eq:opbasis}
\begin{split}
&\mathcal{O}^{(\prime)}_9 = \big[\bar{s}\gamma^\mu P_{L(R)}b \big]\big[\ell\gamma_\mu\ell \big]\, 
 ,\quad \mathcal{O}^{(\prime)}_{10} = \big[\bar{s}\gamma^\mu P_{L(R)}b \big]\big[\ell\gamma_\mu\gamma_5\ell \big]\;,\\
&\mathcal{O}_S^{(\prime)} = \big[\bar{s}P_{R(L)}b \big]\big[\ell \ell \big]\, ,\quad \quad \quad \mathcal{O}_P^{(\prime)} = \big[\bar{s}P_{R(L)}b \big]\big[\ell \gamma_5 \ell \big]
\,.
\end{split}
\end{eqnarray}
    Here, $P_{L,R}=(1\mp\gamma_5)/2$ stand for the (left, right) handed projection operators and the operators $\mathcal{O}_{9(10), S(P)}$ represent the (axial)vector and (pseudo)scalar operators, respectively. The primed operators $\mathcal{O}_i^{\prime}$ can be obtained by flipping the chirality of the operators $\mathcal{O}_i$. The $C^{(\prime)}_{9, 10, S, P}$  are the Wilson coefficients that have zero value in the SM and can have a non-zero value in various new physics scenarios.
%%%%%%%%%%%%%%%%%%%%%%%%%%%%%%
\subsubsection{$b \to s \nu \nu$}
The generic effective Hamiltonian for $b\to s \nu \bar{\nu}$ is as follows~\cite{Buras:2014fpa, Altmannshofer:2009ma}
\begin{align}\label{eq:b2snunu}
    \mathcal{L}_\mathrm{eff}^{b\to s\nu\bar\nu}=\frac{4G_F}{\sqrt{2}}V_{tb}V_{ts}^\ast\frac{\alpha_{em}}{4\pi}\sum_\ell\left(C^{\nu_\ell}_{L}\mathcal{O}^{\nu_\ell}_{L}+C^{\nu_\ell}_{R}\mathcal{O}^{\nu_\ell}_{R}\right)+\mathrm{h.c.}\,,
\end{align}
with
\begin{align}
\mathcal{O}^{\nu_\ell}_{L}=&(\bar{s}\gamma_\mu P_L b)(\bar{\nu}_\ell\gamma^\mu (1-\gamma_5)\nu_\ell)\,, & \mathcal{O}^{\nu_\ell}_{R}=&(\bar{s}\gamma_\mu P_R b)(\bar{\nu}_\ell\gamma^\mu (1-\gamma_5)\nu_\ell)\,.
\end{align}
In Eq.~\eqref{eq:b2snunu},  $V_{tb}$ and $V_{ts}$ are the relevant CKM matrix elements, and the operator $\mathcal{O}_{R}^{\nu_\ell}$ can only arise in the presence of NP, so $C^{\nu_\ell}_{R,\mathrm{SM}}=0$, while $\mathcal{O}_{L}^{\nu_\ell}$ can originate from either the SM or NP, with $C^{\nu_\ell}_{L,\mathrm{SM}}=-6.32(7)$.
%%%%%%%%%%%%%%%%%%%%
\subsection{Standard Model Effective Field Theory}\label{subsec:SMEFT}
In the SMEFT expansion to the dimension-six order, we study the correlations among various low-energy observables. We detail the relevant Lagrangian in the form of the SMEFT operators and then explore various observables that are correlated with $b \to c \ell \nu$ decays in this context.
Above the EW scale, we use the following SMEFT effective Lagrangian at mass dimension six to parameterize model-independent effects of high-scale NP,
\begin{equation}
\mathcal{L_{\mathrm{eff}}} = \mathcal{L_{\mathrm{SM}}} + \sum_{Q_i=Q_i^\dagger} \frac{C_i}{\Lambda^2} Q_i + \sum_{Q_i\neq Q_i^\dagger} \left( \frac{C_i}{\Lambda^2} Q_i + \frac{C_i^\ast}{\Lambda^2} Q_i^\dagger \right) \;, 
\label{eq:SMEFT}
\end{equation}
with the cut-off scale $\Lambda$.

In the SMEFT, the operators which will be relevant for the $b \to c \ell^- \bar{\nu}_{\ell}$ transitions are given by \cite{Aebischer:2015fzz}
\begin{eqnarray} \label{eq:opssmeft}
Q^{(3)}_{\ell q} &=& (\bar{\ell}_i \gamma_{\mu} \tau^{I} \ell_j)(\bar{q}_k \gamma^{\mu} \tau^{I} q_l), ~~~~~~ Q_{\phi u d} = i (\tilde{\phi}^{\dagger} D_{\mu} \phi)(\bar{u}_i \gamma^{\mu} d_j) ,\nonumber \\
Q_{\ell edq} &=& (\bar{\ell}_i^a e_j)(\bar{d}_k q_l^a), ~~~~~~~~~~~~~~~~~ Q^{(1)}_{\ell equ} = (\bar{\ell}_i^a e_j)\epsilon_{ab}(\bar{q}_k^b u_l),\nonumber \\
Q^{(3)}_{\ell equ} &=& (\bar{\ell}_i^a \sigma^{\mu \nu} e_j)\epsilon_{ab}(\bar{q}_k^b \sigma_{\mu \nu} u_l), ~~~~Q^{(3)}_{\phi q} = (\phi^{\dagger} i \overset{\leftrightarrow}{D}_{\mu} \phi)(\bar{q}_i \tau^I \gamma^{\mu} q_j).
\end{eqnarray}
In the above equation, $\ell,q$ and $\phi$ represent lepton, quark and Higgs $SU(2)_L$ doublets, while the right-handed isospin singlets are denoted by $e$, $u$ and $d$.
Here we collect the dimension six operators contributing to the $b \to c \ell \nu$, $b \to s \ell\ell$ and $b \to s \nu\nu$ transitions. Out of the relevant operators, the $Q_{\phi q}^{(3)}$ and $Q _{\phi u d}$ are responsible for the modification of left and right-handed $W$ boson couplings with quarks and leptons, respectively. In this analysis, we focus only on the operators $Q_{lq}^{(3)}$, $ Q_{ledq}$, $Q_{lequ}^{(1)}$ and $Q_{lequ}^{(3)}$ contributing to $b \to c \ell \nu$ transitions as contact interactions.

Now following the method discussed above, a tree-level matching of the SMEFT operators given in Eq.~(\ref{eq:opssmeft}) to the effective operator basis defined in Eq.~ (\ref{eq:effopr}) will result in the following WCs at the $m_b$ scale for $b\to c\ell^-\bar{\nu}_{\ell}$ decays \cite{Aebischer:2015fzz}
\begin{eqnarray} \label{eq:matchedbtoc}
C_{V_L}  &=& -\frac{v^2}{\Lambda^2} \frac{V_{cs}}{V_{cb}} \Big (C^{(3)ll23}_{\ell q}-C_{\phi q}^{(3)} \Big), ~~~~~~~ C_{V_R}   = \frac{v^2}{2\Lambda^2 V_{cb}}C_{\phi u d}, \nonumber \\
C_{S_L}  &=& - \frac{v^2}{2 \Lambda^2} \frac{V_{tb}}{V_{cb}} C^{(1)* ll32}_{\ell equ}, ~~~~~ C_{S_R}   = - \frac{v^2}{2 \Lambda^2} \frac{V_{cs}}{V_{cb}}  C^{* ll32}_{\ell edq}, \nonumber \\
C_T   &=& - \frac{v^2}{2 \Lambda^2} \frac{V_{tb}}{V_{cb}}  C^{*(3) ll32}_{\ell equ} .
\end{eqnarray}
\section{The decay Observables of $B_b \to B_c \ell \bar{\nu}$ processes}
\label{sec:phen}
%%%%%%%%%%%%%%%%%%%%
The differential decay distribution for $B_b \to B_c \ell \nu$ decays, where $B_b$ and $B_c$ represent the bottom and charmed baryons, can be expressed in terms of the helicity amplitudes $H^{V/A}_{\lambda_2\lambda_W}$, as described in Ref.~\cite{Dutta:2013qaa}. This formulation is represented through the parameters $\mathcal{A}_1$, $\mathcal{A}_2$, $\mathcal{A}_3$, and $\mathcal{A}_4$, and is given by:
\begin{equation}
\label{dslnutheta}
\frac{d^2\Gamma}{dq^2\,d\cos\theta} = \mathcal{N}\left(1-\frac{m_\ell^2}{q^2}\right)^2\left[\mathcal{A}_1+\frac{m_\ell^2}{q^2}\mathcal{A}_2+2\mathcal{A}_3+\frac{4m_\ell}{\sqrt{q^2}}\mathcal{A}_4\right],
\end{equation}
where $\theta$ denotes the angle between the momentum vector $\vec{P}_{B_c}$ and the lepton’s three-momentum in the $\ell-\nu$ rest frame. The normalization factor $\mathcal{N}$ is expressed as
\begin{equation}
\mathcal{N} = \frac{G_F^2\,|V_{cb}|^2\,q^2|\vec{P}_{B_c}|}{512\,\pi^3\,m_{B_b}^2}.
\end{equation}
The components $\mathcal{A}_i$ are defined as follows:
\begin{align}
\mathcal{A}_1 &= 2\sin^2\theta\left(H^2_{\frac{1}{2}0}+H^2_{-\frac{1}{2}0}\right)+\left(1-\cos\theta\right)^2H^2_{\frac{1}{2}1}+\left(1+\cos\theta\right)^2H^2_{-\frac{1}{2}-1}, \nonumber \\
\mathcal{A}_2 &= 2\cos^2\theta\left(H^2_{\frac{1}{2}0}+H^2_{-\frac{1}{2}0}\right)+\sin^2\theta\left(H^2_{\frac{1}{2}1}+H^2_{-\frac{1}{2}-1}\right)+2\left(H^2_{\frac{1}{2}t}+H^2_{-\frac{1}{2}t}\right) \nonumber \\
&\quad -4\cos\theta\left(H_{\frac{1}{2}t}H_{\frac{1}{2}0}+H_{-\frac{1}{2}t}H_{-\frac{1}{2}0}\right), \nonumber \\
\mathcal{A}_3 &= \left(H^{SP}_{\frac{1}{2}0}\right)^2 + \left(H^{SP}_{-\frac{1}{2}0}\right)^2, \nonumber \\
\mathcal{A}_4 &= -\cos\theta\left(H_{\frac{1}{2}0}H^{SP}_{\frac{1}{2}0}+H_{-\frac{1}{2}0}H^{SP}_{-\frac{1}{2}0}\right)+ \left(H_{\frac{1}{2}t}H^{SP}_{\frac{1}{2}0}+H_{-\frac{1}{2}t}H^{SP}_{-\frac{1}{2}0}\right).
\end{align} 
The helicity amplitudes $H^{V/A}_{\lambda_2\lambda_W}$ and $H^{SP}_{\lambda_2\lambda_W}$ are essential for defining these terms, and their details are provided in Refs.~\cite{Rajeev:2019ktp, Dutta:2013qaa}. This framework serves as the basis for analyzing the decay observables associated with the process. Now, we define several $q^2$-dependent observables to characterize the decay processes. These include the total differential branching ratio $dB/dq^2$, the ratio of branching fractions $R^{(*)}_{B_c}(q^2)$, the forward-backward asymmetry $A_{FB}^l(q^2)$ and the polarization fraction of the charged lepton $P^{l}(q^2)$. Another key observable is the convexity parameter $C_{F}^l(q^2)$, determined by integrating the $\cos^2\theta$ dependence of the angular distribution. These observables are defined for decay modes $B_b \to B_{c}^{(*)} \ell \bar{\nu}$ as follows:

%%%%%%%%%%%%%%%%%%%%%%
\begin{itemize}
    \item The differential branching ratio
\end{itemize}
\begin{equation}
    d{\rm Br}/dq^2 =\tau_{B_{b}}d\Gamma/dq^2
\end{equation}
%%%%%%%%%%%%%

\begin{itemize}
    \item Lepton non-universal observable
\end{itemize}
\begin{equation}
R_{B_{c}^{(*)}}(q^2)= \frac{\frac{d\Gamma}{dq^2}(B_{b}\to B_{c}^{(*)} \tau^{-}\bar{\nu}_{\tau})}{\frac{d\Gamma}{dq^2}(B_{b}\to B_{c}^{(*)} \ell^{-}\bar{\nu}_{\ell})}
\end{equation}
%%%%%%%%%%%%%%%
\begin{itemize}
    \item Forward-backward asymmetry
\end{itemize}
\begin{equation}
    A_{FB}^{\tau}(q^2)=\frac{\large(\int_{0}^{1}-\int_{1}^{0}\large)\frac{d^2\Gamma}{dq^2 d\cos\theta}d\cos\theta}{\large(\int_{0}^{1}+\int_{1}^{0}\large)\frac{d^2\Gamma}{dq^2 d\cos\theta}d\cos\theta}
\end{equation}
%%%%%%%%%%%%%%%%%%%
\begin{itemize}
    \item Convexity parameter
\end{itemize}
\begin{equation}
     C_{F}^{\tau}(q^2)=\frac{1}{d\Gamma/dq^2}\frac{d^2}{d \cos\theta}\left( \frac{d^2\Gamma}{dq^2d\cos\theta}\right)
\end{equation}.
%%%%%%%%%%%%%%%
\begin{itemize}
    \item Longitudinal polarisation fraction
\end{itemize}
\begin{equation}
   P_{L}^{\tau}(q^2)= \frac{{d\Gamma^{\lambda_\tau=1/2}}/{dq^2}-{d\Gamma^{\lambda_\tau=-1/2}}/{dq^2}}{{d\Gamma^{\lambda_\tau=1/2}}/{dq^2}+{d\Gamma^{\lambda_\tau=-1/2}}/{dq^2}},
\end{equation}
where $d\Gamma^{\lambda_{\tau}=\pm1/2}/dq^2$ are the helicity dependent differential decay rates.

\section{New Physics Analysis in SMEFT}
\label{new physics}
\subsection{Input parameters}
%%%%%%%%%%%%%%%%%%%%
For our analysis, we use various input parameters such as the mass of leptons, mesons, CKM matrix elements, and Fermi coupling constant $G_F$ from PDG~\cite{ParticleDataGroup:2024cfk}. The form factors used in this work are given as follows.

%%%%%%%%%%%%%%%%%%%%%%%%%%%%%%%%%
%%%%%%%%%%%%%%%%%%%%%%%%%%%%%%%
\subsection{Hadronic matrix element and form factor}

For $B_b \to
B_c^{(*)}\ell^{-}\bar{\nu_{\ell}}$ decays, the hadronic matrix elements for vector and axial-vector current can be parametrized in terms of various form factors which are expressed as functions of velocities of baryons~\cite{Wei:2009np}.
For the $\Xi_b\to \Xi_c$ and $\Sigma_b\to\Sigma_c$ transition, the parametrization is as follows:
\begin{eqnarray}
  \label{eq:llff}
  M_{V}^{\mu}=
  \langle B_c(v',s')|V^\mu|B_b(v,s)\rangle&=&\bar
  u_{B_c}(v',s')\Bigl[F_1(\omega)\gamma^\mu+F_2(\omega)v^\mu+F_3(\omega)v'^\mu\Bigl]
u_{B_b}(v,s),\cr
M_{A}^{\mu}= \langle B_c(v',s')|A^\mu|B_b(v,s)\rangle&=& \bar
  u_{B_c}(v',s')\Bigl[G_1(\omega)\gamma^\mu+G_2(\omega)v^\mu+G_3(\omega)v'^\mu\Bigl]
\gamma_5 u_{B_b}(v,s).\qquad 
\end{eqnarray}
The form factors $F_1,F_2,F_3,G_1,G_2,G_3$ are the functions of the velocity transfer variable  
$\omega=v\cdot v'=(m_{B_b}^2+m_{B_c}^2-q^2)/(2m_{B_b}m_{B_c})$, where $v$ and $v'$ are the four-velocity of the baryons $B_b$ and $B_c$. Similarly, the hadronic matrix elements associated with the scalar and pseudoscalar 
currents are given below:
\begin{eqnarray}
\label{eq:llffs}
\langle B_c(v',s') | \bar{c}\,b | B_b(v,s)\rangle &=& \bar{u}_{B_c}(v',s') \Big[
F_1(\omega)\,\frac{\not q}{m_b - m_c} + 
F_2(\omega)\,\left(\frac{m_{B_{b}^{2}}-m_{B_{c}^{2}}+q^2}{2m_{B_b}(m_b - m_c)}\right) \cr
&&+ F_3(\omega)\,\left(\frac{m_{B_{b}^{2}}-m_{B_{c}^{2}}-q^2}{2m_{B_b}(m_b - m_c)}\right) \Big] u_{B_b}(v,s), \cr
\langle B_c(v',s') | \bar{c}\,\gamma_5\,b | B_b(v,s)\rangle &=& \bar{u}_{B_c}(v',s') \Big[
-G_1(\omega)\,\frac{\not q}{m_b + m_c} - 
G_2(w)\,\left(\frac{m_{B_{b}^{2}}-m_{B_{c}^{2}}+q^2}{2m_{B_b}(m_b + m_c)}\right) \cr
&&+ G_3(\omega)\,\left(\frac{m_{B_{b}^{2}}-m_{B_{c}^{2}}-q^2}{2m_{B_b}(m_b + m_c)} \right)\Big]\,\gamma_5\,u_{B_b}(v,s).
\end{eqnarray}
Here $m_b$ and $m_c$ represent the mass of b and c quarks, respectively.

In the heavy quark limit, the form factors for semileptonic decay of $\Xi_b \to \Xi_c$ process reduced to a small number of Isgur-Wise (IW)  functions. This IW function encodes the dynamics of light diquark degrees of freedom. In quasipotential approaches within the quark-diquark picture of heavy baryons, the two light quarks form a diquark state which is described by a diquark wavefunction $\Psi_d$, and the heavy baryon is then considered as a bound state of a heavy quark and light diquark. The heavy quark-diquark baryon wave function is described by the baryon wave function $\Psi_B$. Both $\Psi_d$ and$\Psi_B$ satisfy the relativistic quasipotential equations of the Schr\"odinger  type \cite{Ebert:2006rp},
\begin{equation}
\left( \frac{b^2(M)}{2\mu_R} - \frac{\textbf{p}^2}{2\mu_R} \right) \Psi_{d,B}(\textbf{p}) = \int \frac{d^3q}{(2\pi)^3} \, V(\textbf{p},\textbf{q};M) \, \Psi_{d,B}(\textbf{q}),
\end{equation}
where  $V(\textbf{p},\textbf{q}; M)$ is the quasi-potential operator of the quark-quark or quark-diquark interaction, $b^2(M)$ is the square of the relative momentum   and $\mu_R$ the relativistic reduced mass of the system. Here,\textbf{ p } and \textbf{q} are the relative momentum of light diquark-heavy quark momentum of the initial and final state baryon. The details of the quasipotential operator can be found in  reference \cite{Ebert:2006rp}.\\ The IW function is obtained from the overlap integral of initial and final state baryon wave functions in the heavy quark limit $m_Q \to \infty$,
\begin{equation}
\zeta(\omega) = \lim_{m_Q \to \infty} \int \frac{d^3p}{(2\pi)^3} \, \Psi_{B_{Q'}}\Big(\textbf{p} + 2 \epsilon_d(p) \, e_\Delta \sqrt{\frac{\omega-1}{\omega+1}} \, \Big) \Psi_{B_Q}(\textbf{p}).
\end{equation}
The sub-leading functions can be obtained  $m_Q \to \infty$:
\begin{equation}
\chi(\omega) = -\,\frac{\omega-1}{\omega+1}
\int \frac{d^3 p}{(2\pi)^3}\;
\Psi_{B_{Q'}}\!\left(\mathbf{p} + 2\,\epsilon_d(p)\,\sqrt{\frac{\omega-1}{\omega+1}}\,\mathbf{e}_\Delta \right)
\;\frac{\bar{\Lambda-\epsilon_d(p)}}{\bar{\Lambda}}\Psi_{B_Q}(\mathbf{p}).
\end{equation}
Here $e_\Delta=\Delta/\sqrt{\Delta^2}$ is the unit vector along the recoil momentum $\Delta = M_{B_{Q'}} v' - M_{B_Q} v$, and $\epsilon_d(p)=\sqrt{\mathbf{p}^2+m_d^2}$ is the diquark energy. The subleading $1/m_Q$ corrections introduce an additional function $\chi(w)$, also computed using these wave functions. Here $\omega = v \cdot v'$ is the product of initial and final baryon four-velocities, 
and $\mathbf{e}_\Delta$ is a unit vector in the momentum transfer direction.
Here, the factor $\bar{\Lambda}-\epsilon_d(p)$ reflects the binding of the light 
degrees of freedom in the heavy baryon. Further details can be found in the reference \cite{Ebert:2006rp}. The form factors are described by the IW functions, and the sub-leading terms are expressed as follows:
%%%%%%%%%%%%%%%%%%%%%%%
\begin{eqnarray}
  \label{eq:ffm}
  F_1(\omega)&=&\zeta(\omega)+\left(\frac{\bar\Lambda}{2m_{B_b}}
+\frac{\bar\Lambda}{2m_{B_c}}\right)\left[2\chi(\omega)
+ \zeta(\omega)\right],\cr
G_1(\omega)&=& \zeta(\omega)+\left(\frac{\bar\Lambda}{2m_Q}
+\frac{\bar\Lambda}{2m_{Q'}}\right)\left[2\chi(\omega)
+\frac{\omega-1}{\omega+1} \zeta(w)\right],\cr
F_2(\omega)&=&-\frac{\bar\Lambda}{2m_{Q'}}\frac{2}{\omega+1}\zeta(\omega),\cr
G_2(\omega)&=&-\frac{\bar\Lambda}{2m_{Q'}}\frac{2}{\omega+1}\zeta(\omega)
,\cr
F_3(\omega)&=&-G_3(\omega)=-\frac{\bar\Lambda}{2m_{Q}}\frac{2}{\omega+1}\zeta(\omega)
,
\end{eqnarray}
where the parameter $\bar{\Lambda}=(m_{B_b}-m_b)$ and $\zeta(\omega)$ is the Isgur-Wise (IW) function. The $\chi(\omega)$ function arises due to the additional $1/m_b$ correction to the HQET Lagrangian. The functions mentioned above are approximated at the zero recoil point and are expressed as follows:
%%%%%%%%%%%
\begin{eqnarray}
  \label{eq:exp}
  \zeta(\omega)&=&1-\rho_{\zeta}^2(\omega-1)+c_{\zeta}(\omega-1)^2+\cdots,\cr
\chi(\omega)&=&\rho_\chi^2(\omega-1)+c_\chi(\omega-1)^2+\cdots.
\end{eqnarray}
The parameter $\rho_{\zeta,\chi}^2$ gives the slope, and $c_{\zeta,\chi}$ gives the curvature of IW functions. The values of these parameters are  $\rho_{\zeta}^2=2.27$, $c_{\zeta}=3.87$, $\rho_{\chi}^2=0.045$ and $c_{\chi}=0.036$ respectively.
\\
%%%%%%%%%%%%%%%%%%%%%%
For $\Sigma_b\to
\Sigma_c^*$  transition the parametrization is given as follows:
\begin{eqnarray}
  \label{eq:llffs}
  \langle B_c^*(v',s')|V^\mu|B_b(v,s)\rangle= \bar
  u_{B_c^*,\lambda }(v',s')\!\!&\Bigl[&\!\! N_1(\omega)v^\lambda
\gamma^\mu+N_2(\omega)v^\lambda v^\mu\cr
&&+N_3(\omega)v^\lambda v'^\mu+ N_4(\omega)
g^{\lambda\mu}\Bigl]\gamma_5 u_{B_b}(v,s),\cr
 \langle B_c^*(v',s')|A^\mu|B_b(v,s)\rangle= \bar
  u_{B_c^*,\lambda}(v',s')\!\!&\Bigl[&\!\!K_1(\omega)v^\lambda\gamma^\mu
+K_2(\omega)v^\lambda v^\mu\cr
&&+K_3(\omega)v^\lambda v'^\mu+ K_4(\omega)g^{\lambda\mu}\Bigl]
u_{B_b}(v,s).
\end{eqnarray}
%%%%%%%%%%%%%%%%%%%%%%%%%

Again, we use the equation of motion to find the hadronic matrix elements of scalar and pseudo-scalar
currents, which are given as follows,
\begin{eqnarray}
\label{eq:llffs_v2}
&&\langle B_c^*(v',s')| \bar{c}\,b | B_b(v,s)\rangle = \bar{u}_{B_c^*}(v',s') \Big[
N_1(w)v^{\lambda}\,\frac{\not q}{m_b - m_c} + 
N_2(w)v^{\lambda}\,\left(\frac{m_{B_{b}^{2}}-m_{B_{c}^{2}}+q^2}{2m_{B_b}(m_b - m_c)}\right) \cr
&&~~~~~~~~~~~~~~~~~~~~~~~~~~+ N_3(\omega)v^{\lambda}\,\left(\frac{m_{B_{b}^{2}}-m_{B_{c}^{2}}-q^2}{2m_{B_b}(m_b - m_c)}\right) -N_4(\omega)\,\frac{\not q}{m_b + m_c}\Big] u_{B_b}(v,s), \cr
&&\langle B_c^*(v',s') | \bar{c}\,\gamma_5\,b | B_b(v,s)\rangle = \bar{u}_{B_c^*}(v',s') \Big[
-K_1(\omega)v^{\lambda}\,\frac{\not q}{m_b + m_c} - 
K_2(\omega)v^{\lambda}\,\left(\frac{m_{B_{b}^{2}}-m_{B_{c}^{2}}+q^2}{2m_{B_b}(m_b + m_c)}\right) \cr
&&~~~~~~~~~~~~~~~~~~~~~~~~~-K_3(\omega)v^{\lambda}\,\left(\frac{m_{B_{b}^{2}}-m_{B_{c}^{2}}-q^2}{2m_{B_b}(m_b + m_c)}\right)-K_4(\omega)\,\frac{ q^{\lambda}}{m_b+m_c}\Big]\,\gamma_5\,u_{B_b}(v,s),
\end{eqnarray}
where $ u_{B_c^*,\lambda }$ denotes the Rarita-Schwinger spinors for the $B_{c}^*$ baryon.

%%%%%%%%%%%%%%%%%%%%%%%%%%

For the semileptonic $\Sigma_b \to \Sigma_{c}^*$ transition, the form factors in the heavy quark limit can be expressed as
%%%%%%%%%%%%%%%
\begin{eqnarray}
  \label{eq:ffo}
  F_1(\omega)&=&G_1(\omega)=-\frac13\zeta_1(\omega),\cr
F_2(\omega)&=&\frac23\frac2{\omega+1}\zeta_1(w),\cr
G_2(\omega)&=&G_3(\omega)=0,\cr
N_1(\omega)&=&-N_3(\omega)=K_3(\omega)=-\frac1{\sqrt3}\frac2{\omega+1}\zeta_1(\omega),\cr
N_4(\omega)&=&-K_4(\omega)=-\frac2{\sqrt3}\zeta_1(\omega),\cr
N_2(\omega)&=&K_1(\omega)=K_2(\omega)=0.
\end{eqnarray}
Here the IW function $\zeta_{1}(\omega)$ can be approximated as
\begin{equation}
  \label{eq:expzeta}
  \zeta_{1}(\omega)=\zeta_1(1)-\rho_{\zeta_1}^2(\omega-1)+c_{\zeta_1}(\omega-1)^2+\cdots,
\end{equation}
with $\rho_{\zeta_1}^2=2.17$ and $c_{\zeta_1}=3.36$.

%%%%%%%%%%%%%%%%%%%%
%%%%%%%%%%%%%%%%%%%%%%
\subsection{Constraints on the SMEFT coefficient(s)}
%%%%%%%%%%%%%
This subsection focuses on evaluating the SMEFT Wilson coefficients, such as $C_{\ell q}^{(3)}$, $C_{\ell e q u}^{(1)}$, and $C_{\ell e d q}$, by analyzing various $B$ mesonic observables mediated by the $b \to c \tau \nu_{\tau}$ transitions. We also examine observables associated with $b \to s \tau^+ \tau^-$ and $b \to s \nu \nu$ transitions to impose complementary constraints on the SMEFT Wilson coefficients. However, among the couplings, only $C_{\ell q}^{(3)}$ contributes to $b \to c \tau \nu$, $b \to s \tau ^+ \tau ^-$ and $b \to s \nu \bar{\nu}$ mediated decays. 

\begin{figure}[h!]
\centering
\includegraphics[height=40mm,width=50mm]{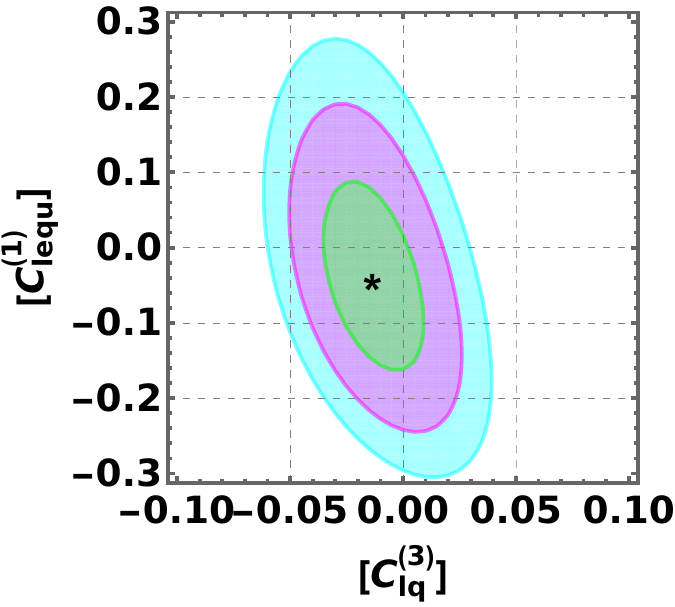}
\hspace{1mm}  % Adds space between the first and second plot
\includegraphics[height=40mm,width=50mm]{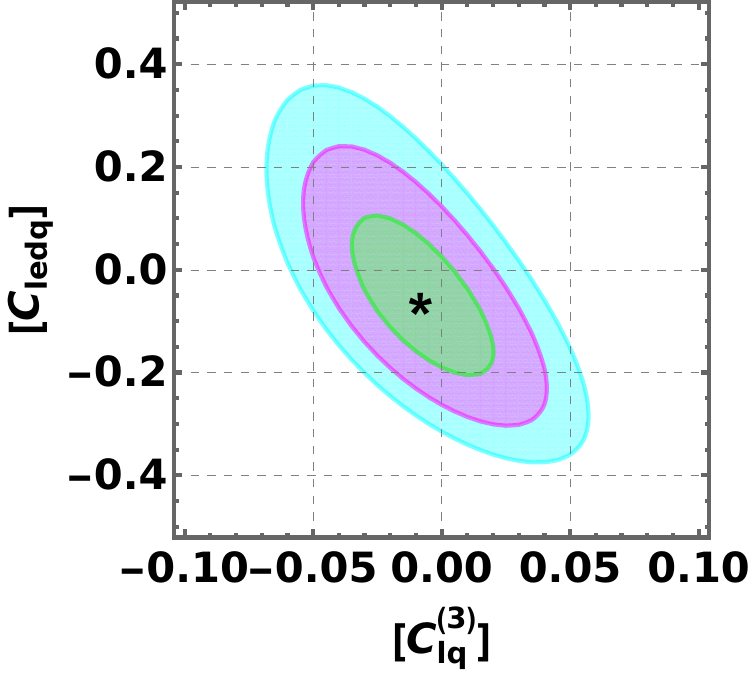}
\hspace{1mm}  % Adds space between the second and third plot
\includegraphics[height=40mm,width=50mm]{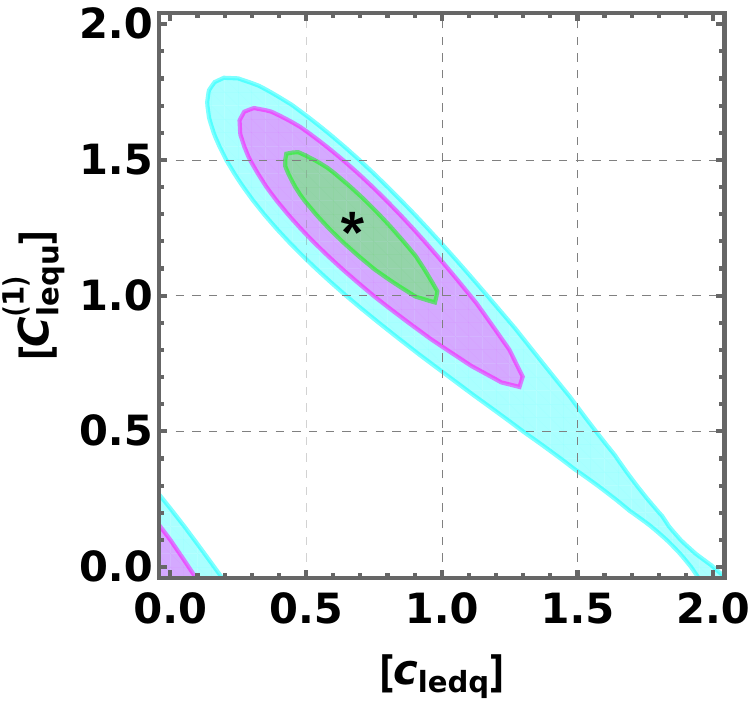}
\caption{Allowed parameter scan in 1$\sigma$ (green), 2$\sigma$ (magenta), and 3$\sigma$ (cyan) regions. The black star denotes the best-fit value.}
\label{fig::parameter space}
\end{figure}

We employ the experimental observations given in Table~\ref{tab:UpperBounds-inputs1} to constrain the NP parameters. The allowed parameter scan of the NP couplings in 1$\sigma$, 2$\sigma$ and 3$\sigma$ regions are presented in Fig. \ref{fig::parameter space} where a naive chi-square analysis has been used to obtain the best fit values.
 
 \begin{table}[htb]
\centering
\begin{tabular}{||c|c||}
\hline \hline
\textbf{Observables} & \textbf{Experimental Value} \\ \hline
\rowcolor{gray!35}$R_{D}$  & $0.357 \pm 0.029 \pm 0.014$~\cite{HFLAV:2022link}\\   \hline
\rowcolor{gray!30}$R_{D^{*}}$  & $0.284 \pm 0.010 \pm 0.012$ ~\cite{HFLAV:2022link}  \\ \hline
\rowcolor{gray!30}$P_{\tau}(D^{*})$ & $-0.38 \pm 0.51^{+0.21}_{-0.16}$~\cite{Belle:2019ewo} \\   \hline
\rowcolor{gray!30}$F_{L}(D^{*})$ & $0.60 \pm 0.08 \pm 0.04$~\cite{Belle:2019ewo} \\   \hline
\rowcolor{gray!30}$R(\Lambda_{c})$ & $0.242 \pm 0.026 \pm \pm 0.040$~\cite{LHCb:2022piu} \\ \hline \hline
\rowcolor{gray!20}$\mathcal{B}(B \to K^{+}\nu \bar{\nu})$ & $(2.3 \pm 0.7) \times 10^{-5}$~\cite{Belle-II:2023esi} \\    \hline
\rowcolor{gray!20}$\mathcal{B}(B^{0} \to K^{*0}\nu \bar{\nu})$ & $<1.8 \times 10^{-5}$~\cite{Belle:2017oht} \\ \hline \hline
\rowcolor{gray!30}$\mathcal{B}(B_s \to \tau^{+}\tau^{-})$ & $<6.8 \times 10^{-3}$~\cite{LHCb:2017myy} \\    \hline
\rowcolor{gray!30}$\mathcal{B}(\bar{B} \to K^{+}\tau^{+} \tau^{-})$ & $<2.25 \times 10^{-3}$~\cite{BaBar:2016wgb} \\ 
\hline \hline
\end{tabular}
\caption{Experimental values of various $b \to c \tau \nu_{\tau}$, $b \to s \nu \bar{\nu}$, and $b \to s \tau^{+} \tau^{-}$ observables.} 
\label{tab:UpperBounds-inputs1}
\end{table}

%%%%%%%%%%%%%%%%%%%%%%%%%%%%%%
 The relevant formula for the chi-square analysis is defined as 
 \bea
\chi^2(C^{\rm NP})= \sum_i  \frac{\Big ({\cal O}_i^{\rm Th}(C^{\rm NP}) -{\cal O}_i^{\rm Exp} \Big )^2}{(\Delta {\cal O}_i^{\rm Exp})^2+(\Delta {\cal O}_i^{\rm SM})^2},
\eea
where ${\cal O}_I ^ {\rm Th}$ and ${\cal O}_I ^ {\rm Exp}$ represent the theoretical values and the measured central value of the observables, respectively. The denominator represents the error associated with the SM and experimental values.
From this analysis, the best-fit values of the SMEFT couplings in 1D and 2D scenarios are provided in Table \ref{tab:UpperBounds-inputs}.

\begin{table}[htb]
\begin{center}
\begin{tabular}{||c|c|c|c||}
\hline \hline
SMEFT couplings (1$\sigma$)~~~~~&$b \to c \tau \nu_{\tau}$&$b \to s \tau^{+} \tau^{-} $&$b \to s \nu \bar{\nu}$ \\
\hline 
\rowcolor{gray!30}$C_{lq}^{(3)}$  &$ 1.400$ $[\substack{1.429 \\1.371 } ]$~~~&$ 0.440$ $[\substack{0.469 \\0.420 } ]$& $0.093$ $[\substack{0.136\\0.044 } ]$~~~~~~~ \\ \hline
\rowcolor{gray!30}$C_{lequ}^{(1)}$  &$ -0.055$ $[\substack{0.110 \\-0.220 } ]$~~~&---&---  \\ \hline
\rowcolor{gray!30}$C_{ledq}$  &$ -0.087$ $[\substack{0.064 \\-0.238 } ]$~&---&---  \\ \hline
\rowcolor{gray!20}$(C_{lq}^{(3)}, C_{lequ}^{(1)})$ &~~~$\big( -0.013$ $[\substack{0.017 \\-0.044 }]$, $ -0.042$ $[\substack{ -0.001\\-0.083 }] \big )$  ~&---&---  \\ \hline
\rowcolor{gray!20}$(C_{lq}^{(3)}, C_{ledq})$  &$\big(-0.007 $$  
 [\substack{ 0.021 \\-0.035}]$, $-0.058$$ [\substack{-0.046 \\ -0.07}]\big)$&---&---  \\ \hline
\rowcolor{gray!20}$(C_{lequ}^{(1)}, C_{ledq})$ &$\big(1.283$$   [\substack{ 1.450 \\1.110}]$, $0.675$$  [\substack{ 0.940\\ 0.360}]\big)$&---&--- \\ 
 \hline \hline
\end{tabular}
\caption{Best-fit values $[1\sigma]$ of NP couplings  in 1D and 2D scenario(s).} 
\label{tab:UpperBounds-inputs}
\end{center}
\end{table}

%%%%%%%%%%%%%%%%%%%%%%%%%%%%%%%%%%%%%%%%%%%%%%%%

% \begin{figure}[h!]
% \centering
% \includegraphics[height=55mm,width=75mm]{b_s_n_n.pdf}
% \hspace{2mm}
% \includegraphics[height=55mm,width=75mm]{b_s_t_t.pdf}
% \caption{Branching ratio (in the units of $10^{-6}$) of  $B \to  K^{*}\tau \mu $ (left) and  $B \to \phi \tau \mu $ (right).}
% \label{fig::BRM}
% \end{figure}
% \label{sec:num}
%%%%%%%%%%%%%%%%%%%%%%%%%%%%
\section{Results and Discussions}
\label{result and discussion}
After determining the SMEFT couplings using the measured observables given in Table \ref{tab:UpperBounds-inputs1}, we analyze the $q^2$-dependence of several observables associated with the baryonic decay modes $\Sigma_b \to \Sigma_c^{(*)} \tau^-\bar{\nu}_\tau$ and $\Xi_b \to \Xi_c \tau^-\bar{\nu}_\tau$. The impact of the three one-dimensional scenarios is found to be less significant; hence, our focus shifts to the two-dimensional scenarios.
The  details of our analysis are outlined below.
%%%%%%%%
\subsection{Analysis of $\Xi_b \to \Xi_c \tau^-\bar{\nu}_\tau$Process}
%%%%%%%%%%%%%%%%%%%%%
\begin{figure}[h!]
\centering
\includegraphics[height=40mm,width=50mm]{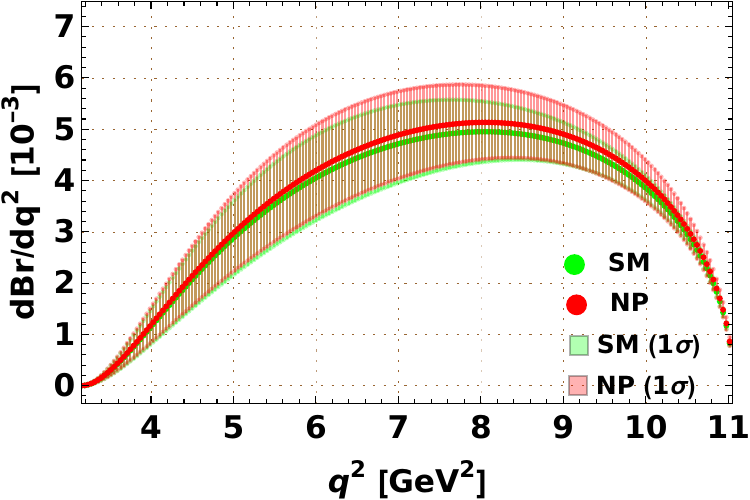}
\hspace{1mm}
\includegraphics[height=40mm,width=50mm]{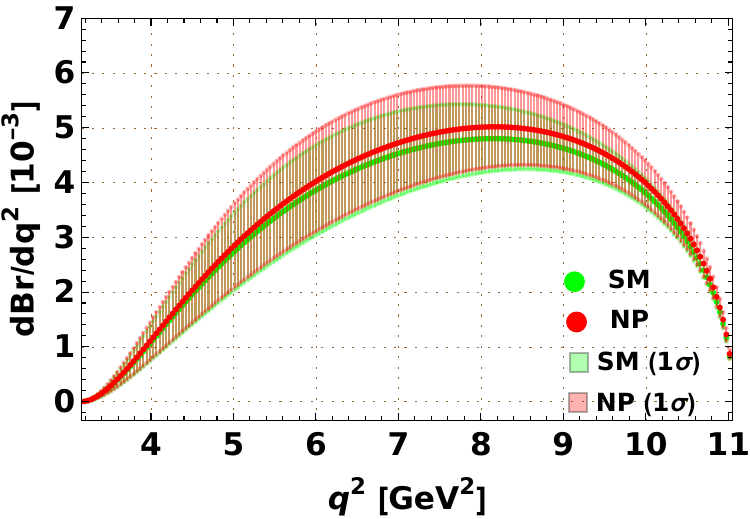}
\hspace{1mm}
\includegraphics[height=40mm,width=50mm]{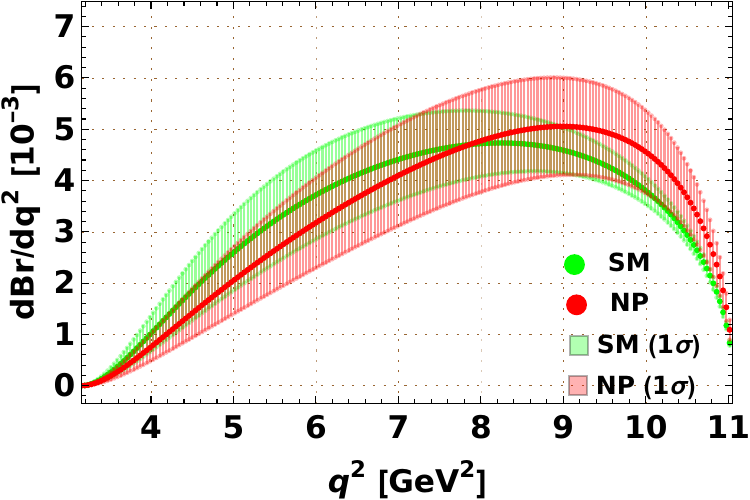}
\hspace{1mm}
\includegraphics[height=40mm,width=50mm]{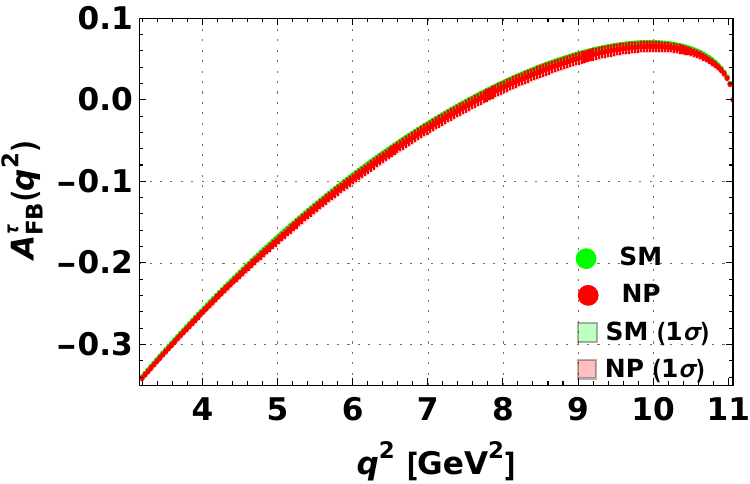}
\hspace{1mm}
\includegraphics[height=40mm,width=50mm]{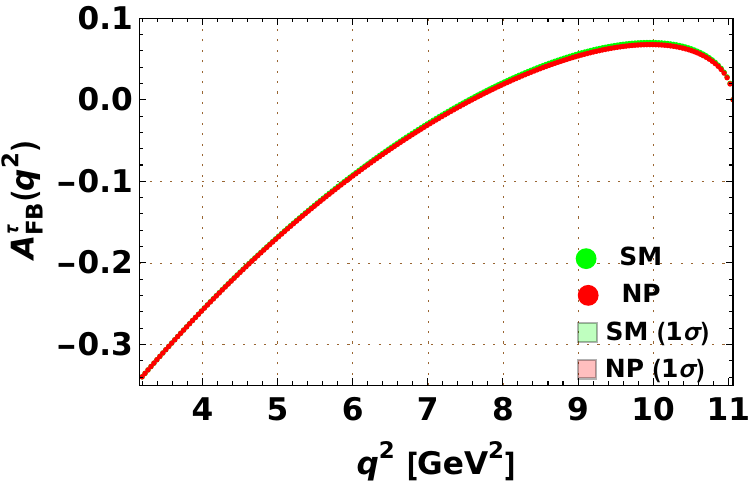}
\hspace{1mm}
\includegraphics[height=40mm,width=50mm]{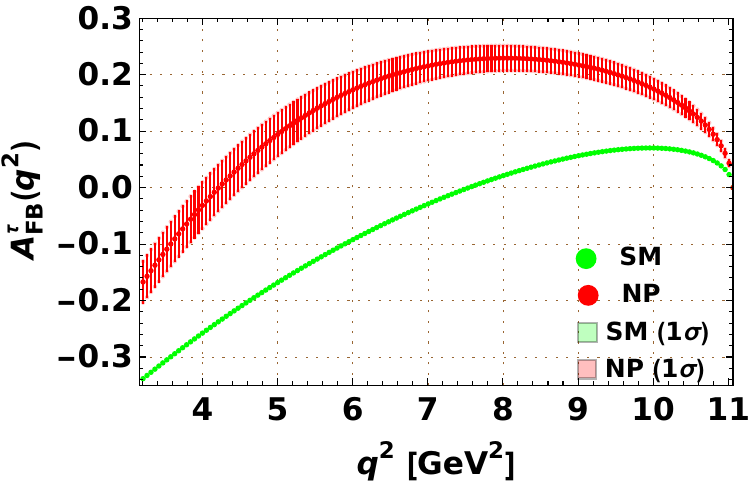}
%%%%%
\caption{$d\text{B}r/dq^2$ and $A_{FB}$ for $\Xi_b \to \Xi_c \tau^- \bar{\nu}_\tau$ in $(C_{lq}^{(3)}, C_{lequ}^{(1)})$ (left), $(C_{lq}^{(3)}, C_{ledq})$ (middle), and $(C_{lequ}^{(1)}, C_{ledq})$ (right). Top panel: branching ratio, Bottom panel: forward-backward asymmetry.}
\label{fig::BRCascade}
\end{figure}
%%%%%%%
%%%%%%%%%%%%%

Fig. \ref{fig::BRCascade} (top panel) illustrates the variation of the differential branching ratio as a function of di-lepton mass squared ($q^2$) distribution. The green band represents the SM contribution, while the red band corresponds to predictions under the new physics framework. Among the three two-dimensional scenarios, the combination of Wilson coefficients $(C_{lequ}^{(1)}, C_{ledq})$ exhibits a significant deviation from the SM, emphasizing the potential impact of the associated operators on the observed behavior.
%%%%%%

Similarly, Fig. \ref{fig::BRCascade} (bottom panel) shows the forward-backward asymmetry, revealing a pronounced shift in the zero-crossing point from approximately $4.2~\mathrm{GeV}^2$ to $7.5~\mathrm{GeV}^2$. This behavior strongly suggests the influence of new physics contributions associated with the Wilson coefficients $(C_{lequ}^{(1)}, C_{ledq})$. In contrast, the other scenarios remain largely consistent with the SM predictions.
%%%%%%%%%

%%%%%%%%%%%
%%%%%%%%%%%%%

%%%%%%%%%%

\begin{figure}[h!]
\centering
\includegraphics[height=40mm,width=50mm]{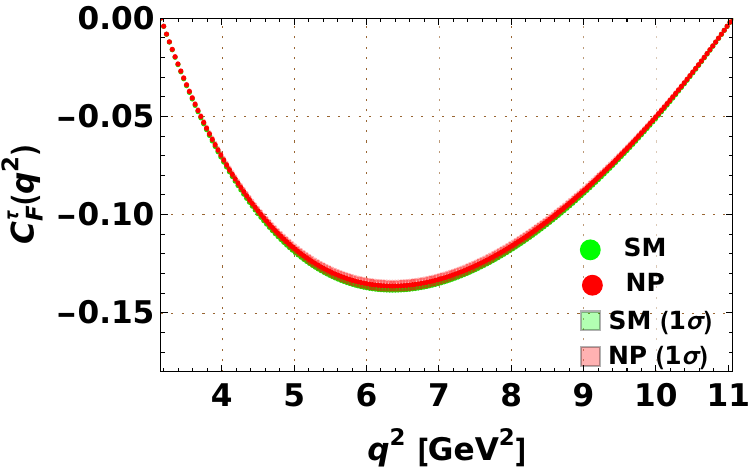}
\quad
\includegraphics[height=40mm,width=50mm]{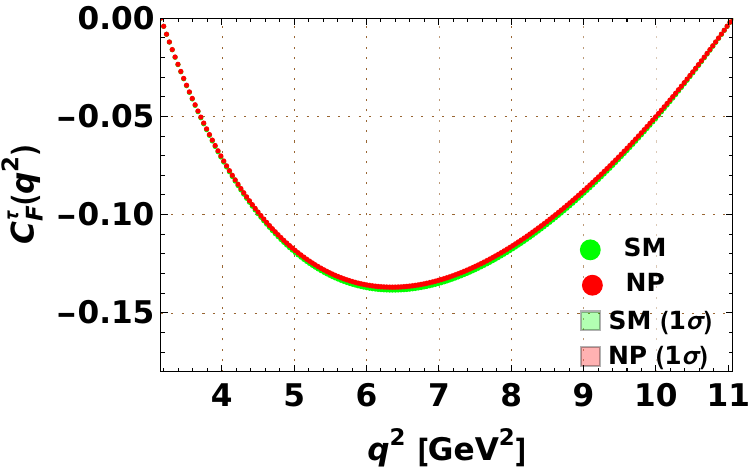}
\includegraphics[height=40mm,width=50mm]{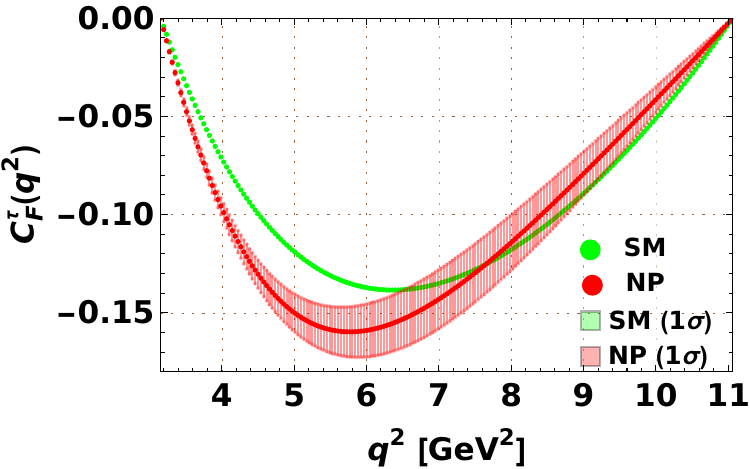}

\caption{The $q^2$ distribution of convexity parameter of $\Xi_b \to \Xi_c \tau^-\bar{\nu}_\tau$ process. The description of couplings is same as Fig. \ref{fig::BRCascade}}
\label{fig::CCascade}
\end{figure}

%%%%%%%%%%%%%%
% \ding{43}
%%%%%%%
%%%%%%%%%%
On the other hand, we observe a notable convexity pattern in the low \( q^2 \) region, characterized by negligible uncertainty, arising from the simultaneous impact of the new physics contributions of the \( C_{lequ}^{(1)} \) and \( C_{ledq} \) couplings. This is shown in Fig. \ref{fig::CCascade}.

\begin{figure}[h!]
\centering
\includegraphics[height=40mm,width=50mm]{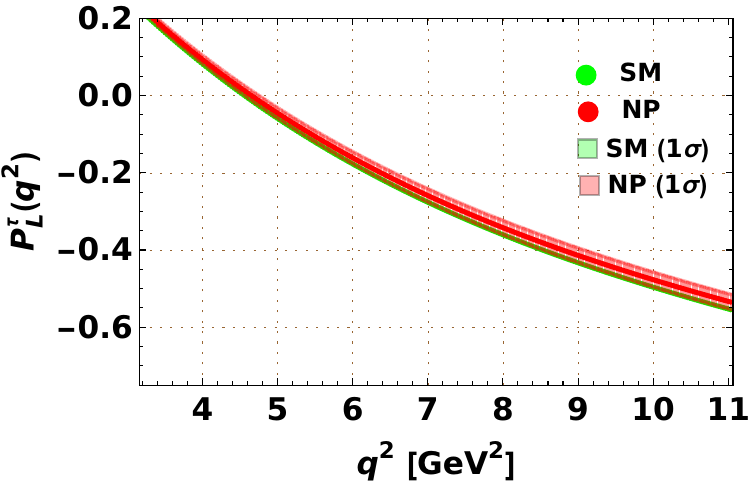}
\includegraphics[height=40mm,width=50mm]{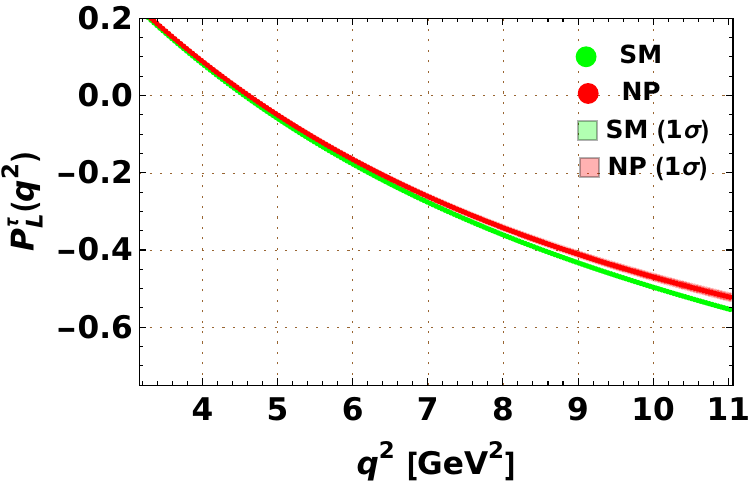}
\includegraphics[height=40mm,width=50mm]{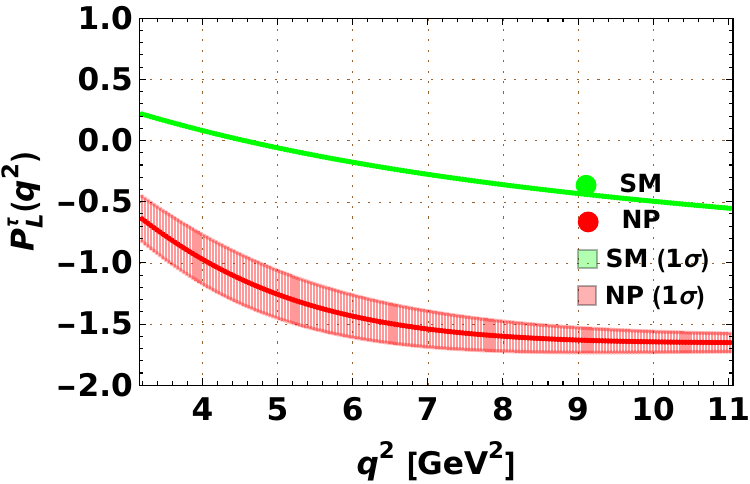}
\caption{The $q^2$ dependence of the longitudinal polarization fraction $\Xi_b \to \Xi_c \tau^-\bar{\nu}_\tau$ process. The description of couplings is same as Fig. \ref{fig::BRCascade}.}
\label{fig::PLCascade}
\end{figure}

%%%%%%%%%%%%%%%%%%%%%
Fig. \ref{fig::PLCascade} depicts the deviation in the lepton polarisation asymmetry observable, which indicates prominent deviation for the entire $q^2$ range in the presence of the couplings $(C_{lequ}^{(1)}, C_{ledq})$. However, we observe a mild deviation the presence of $(C_{lq}^{(3)},C_{ledq})$ in the high $q^2$ region.
%%%%%%%%
%%%%%%%%%%%%%

\begin{figure}[h!]
\centering
\includegraphics[height=40mm,width=50mm]{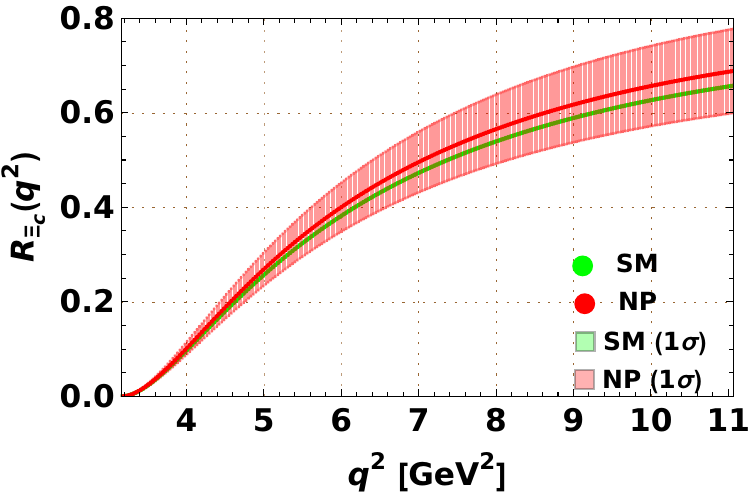}
\quad
\includegraphics[height=40mm,width=50mm]{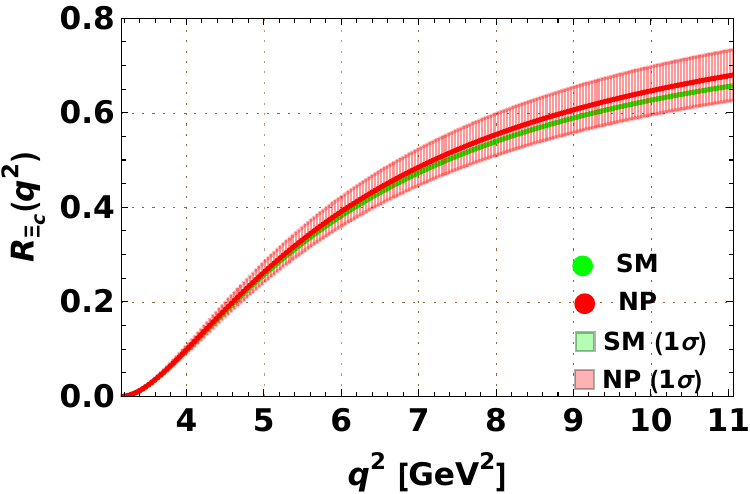}
\includegraphics[height=40mm,width=50mm]{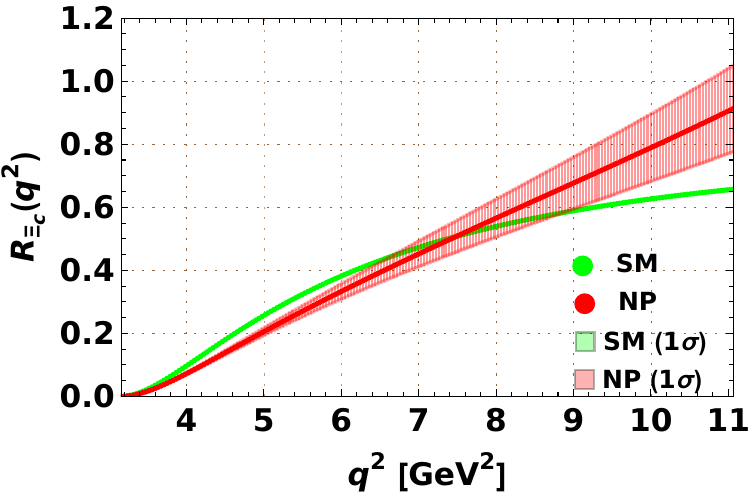}
\caption{The $q^2$ behavior of $R_{\Sigma_b}$ in the $\Xi_b \to \Xi_c \tau^-\bar{\nu}_\tau$ process. The description of couplings is same as Fig. \ref{fig::BRCascade}.}
\label{fig::RCascade}
\end{figure}

We also analyze the $q^2$-dependence of the lepton non-universality observable, as illustrated in Fig.~\ref{fig::RCascade}. A significant deviation from the SM prediction is observed at large $q^2$ values, particularly evident in the right panel. In contrast, the low $q^2$ region shows only a mild discrepancy from the SM. Notably, such discrepancies are absent in the other two scenarios, where the predictions align closely with the SM across the entire $q^2$ range. This indicates that the observed deviations are strongly dependent on the specific scenario considered.

%%%%%%%%%%%%%

%%%%%%%

%%%%%%%%%%

% \paragraph{Comments on complementary constraints}
 % Due to $SU(2)_{L}$ relations, the operators enter the leptonic and semileptonic B decays with underlying quark level transition $b\to s\nu\nu$ and $b \to s\ell\ell $transition.hence we incorporate the observables to provide bounds on the WCs Specifically the scenario in which only $C_{lq}^{(3)}$ and $C_{\phi q}^{(3)}$ is present. considering $b \to s \nu\nu $process the result matches the SM to the scenario in case of $b \to c \ell\nu$ process. whereas bound obtained from the $b\to s \tau\tau $ process incorporating this fits the NP differential branching ratio curves shows significant deviation from SM prediction. Due to the difficulty in tau reconstruction, the upper limit on the branching ratio on the decay is far beyond the order of the SM prediction. therefore it explores a larger parameter space compared to the $b\to s \nu\nu$ and $b \to c\ell\nu$ process.
%%%%%%%%%%%%%%%%%%%%%%%%%%%%%%%%%%%%%
\subsection{Analysis of  $\Sigma_b \to \Sigma_c \tau^-\bar{\nu}_\tau$ Process}
%%%%%%%%%%%%
This subsection will explore the various observables of the $\Sigma_b \to \Sigma_c \tau^-\bar{\nu}_\tau$ channel. The behavior of the $q^2$ distribution are discussed below.
%%%%%%%%%
\begin{figure}[h!]
\centering
\includegraphics[height=40mm,width=50mm]{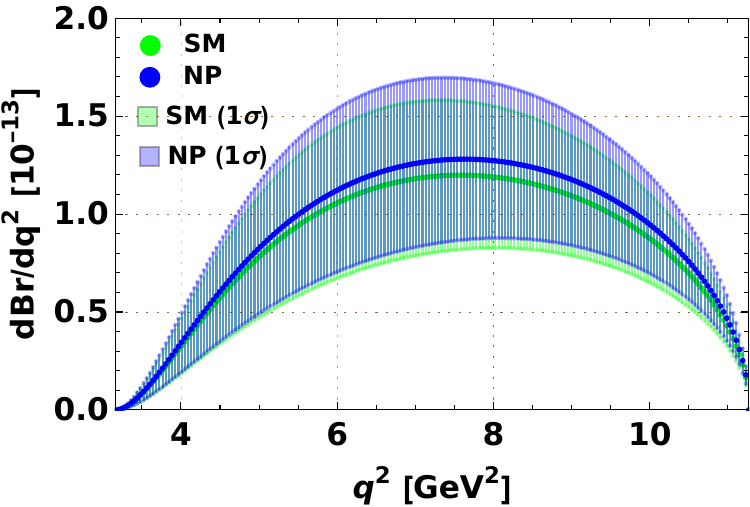}
\includegraphics[height=40mm,width=50mm]{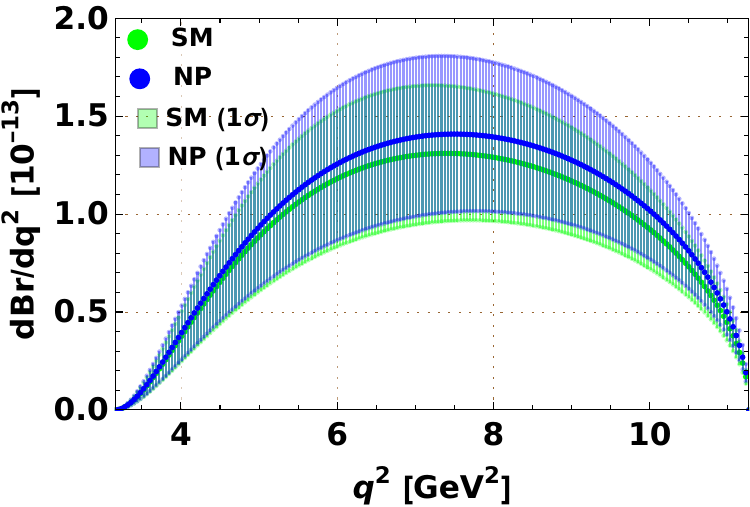}
\includegraphics[height=40mm,width=50mm]{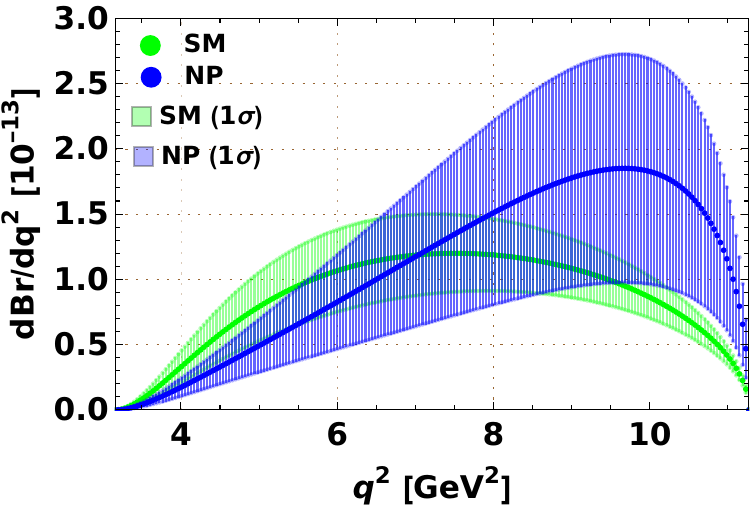}
\quad
\includegraphics[height=40mm,width=50mm]{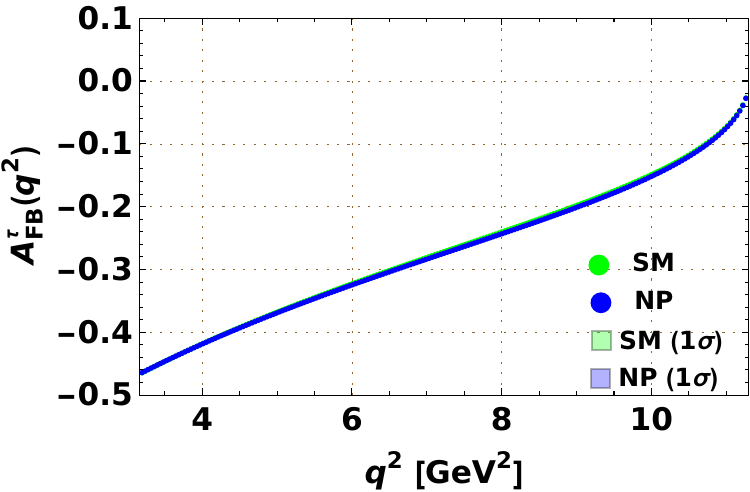}
\includegraphics[height=40mm,width=50mm]{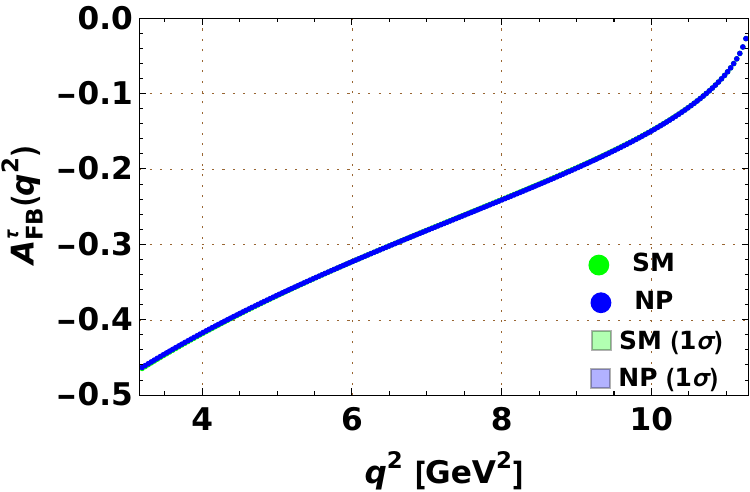}
\includegraphics[height=40mm,width=50mm]{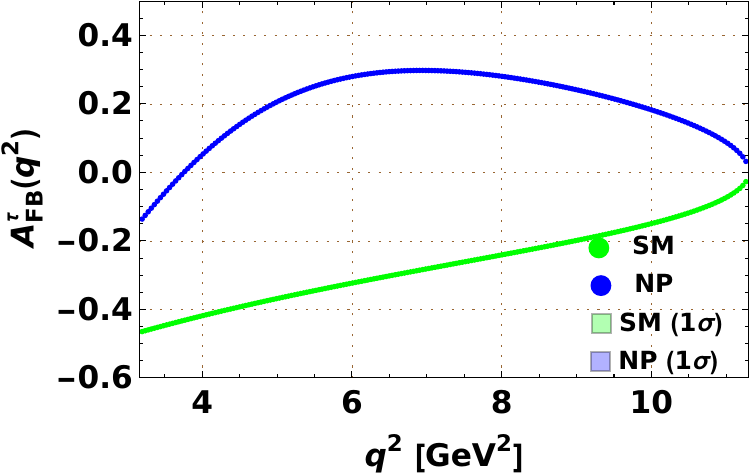}
\caption{$d\text{B}r/dq^2$ and $A_{FB}$ for $\Sigma_b \to \Sigma_c \tau^- \bar{\nu}_\tau$ processes. The description of couplings is same as Fig. \ref{fig::BRCascade}. Top panel: branching ratio, Bottom panel: forward-backward asymmetry.}
\label{fig::BRSigma}
\end{figure}

%%%%%%%%%
%\ding{43} \textbf{Branching ratio}:
Figure~\ref{fig::BRSigma} (top panel) depicts the variation of the differential branching ratio as a function of \( q^2 \). The green band represents the prediction based on the SM contribution. In contrast, the blue band corresponds to the predictions obtained within the framework of SMEFT, incorporating the effects of new physics couplings. Among the three considered two-parameter scenarios, the combination $(C_{lequ}^{(1)}, C_{ledq})$ shows a pronounced deviation from the SM prediction across the higher  $q^2$ spectrum. This notable divergence suggests that these particular operators introduce significant contributions, which may point to the underlying new physics effects.
%%%%%%
% \begin{figure}[h!]
% \centering
% % \includegraphics[height=55mm,width=70mm]{S_CVL_AFB.pdf}
% \includegraphics[height=40mm,width=50mm]{S_AFB_CVL_CSL.pdf}
% \includegraphics[height=40mm,width=50mm]{S_AFB_CVL_CSR.pdf}
% \includegraphics[height=40mm,width=50mm]{S_AFB_CSL_CSR.pdf}
% \caption{$q^2$ dependent Forward backward asymmetry of $\Sigma_b \to \Sigma_c \tau^-\bar{\nu}_\tau$ in $(C_{lq}^{(3)}, C_{\phi q}^{(3)})$(Top left) $(C_{lq}^{(3)},C_{lequ}^{(1)})$(Top right),$(C_{lq}^{(3)}=-C_{\phi q}^{(3)},C_{ledq})$(Bottom left) and $(C_{lequ}^{(1)}, C_{ledq})$(bottom right).}
% \label{fig::AFBSigma}
% \end{figure}
%%%%%%%%

Considering the lepton forward-backward asymmetry, the decay $\Sigma_b \to \Sigma_c \tau \nu$ demonstrates a characteristic zero crossing in the FBA curve within the SM. However, in the presence of new physics contributions from the Wilson coefficients  $(C_{lequ}^{(1)}, C_{ledq})$, the behavior of the observable remains negative across the entire $q^2$ range. This behavior is illustrated in Fig.~\ref{fig::BRSigma} (bottom panel). In contrast, for the other two 2D scenarios, $(C_{lq}^{(3)}, C_{\phi q}^{(3)})$  and $(C_{lq}^{(3)}, C_{lequ}^{(1)})$, the zero crossing of the FBA curve persists, closely resembling the SM prediction.

 %%%%%%%
\begin{figure}[h!]
\centering
\includegraphics[height=40mm,width=50mm]{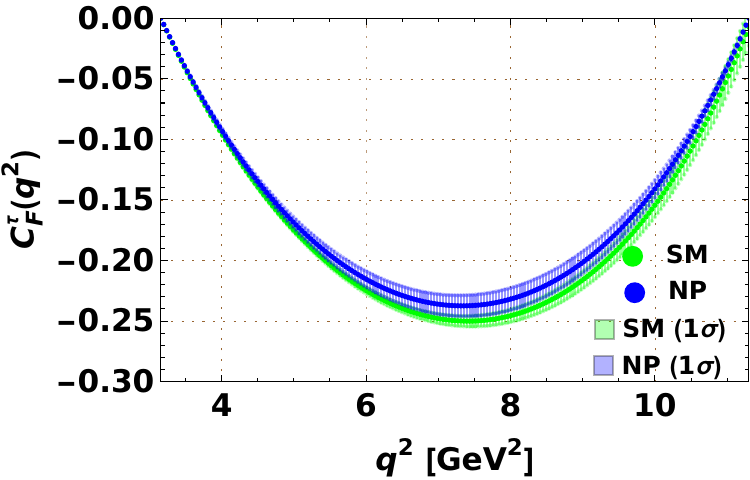}
\includegraphics[height=40mm,width=50mm]{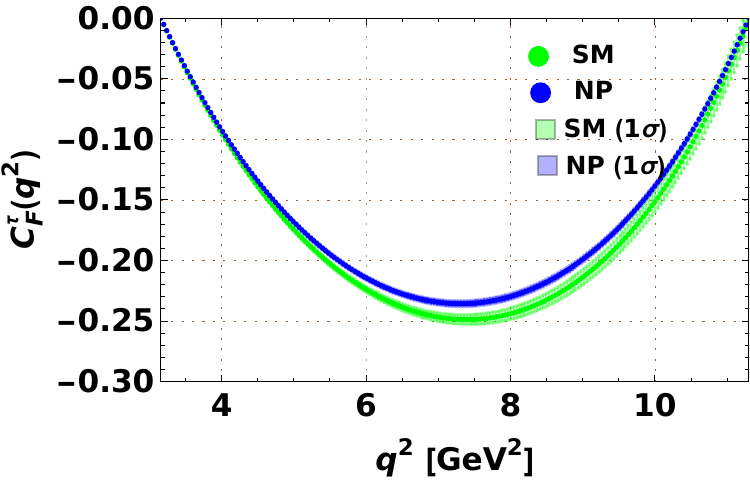}
\includegraphics[height=40mm,width=50mm]{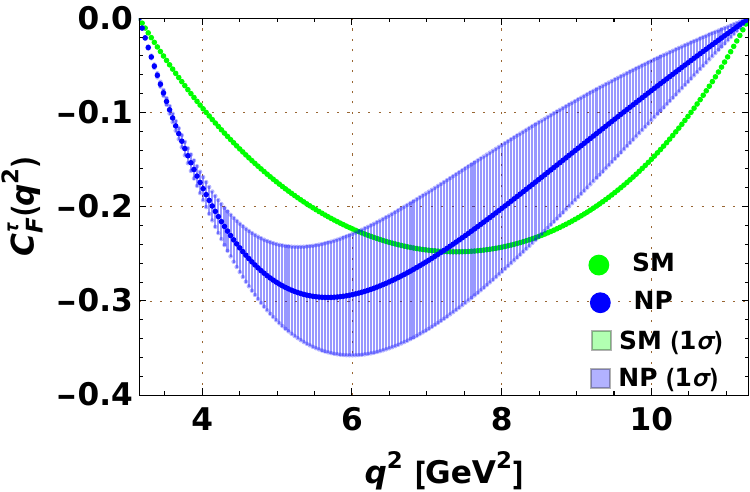}
\caption{$q^2$ dependence of the convexity parameter of $\Sigma_b \to \Sigma_c \tau^-\bar{\nu}_\tau$ process. The description of couplings is same as Fig. \ref{fig::BRCascade}.}
\label{fig::CSigma}
\end{figure}
\begin{figure}[h!]
\centering
\includegraphics[height=40mm,width=50mm]{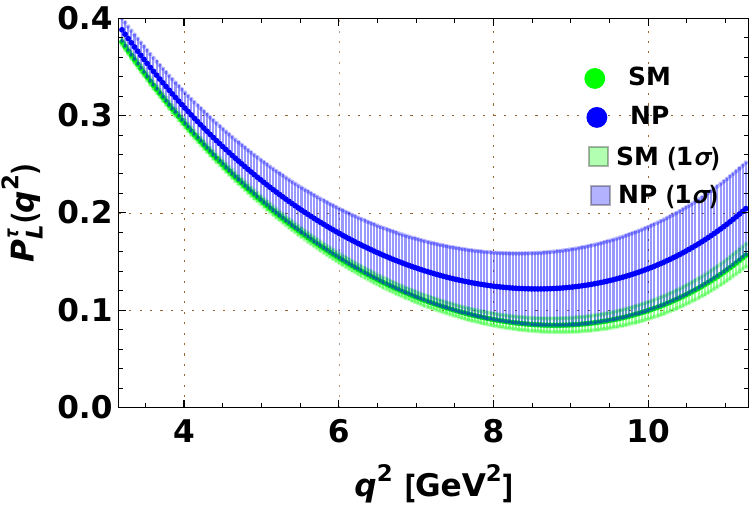}
\includegraphics[height=40mm,width=50mm]{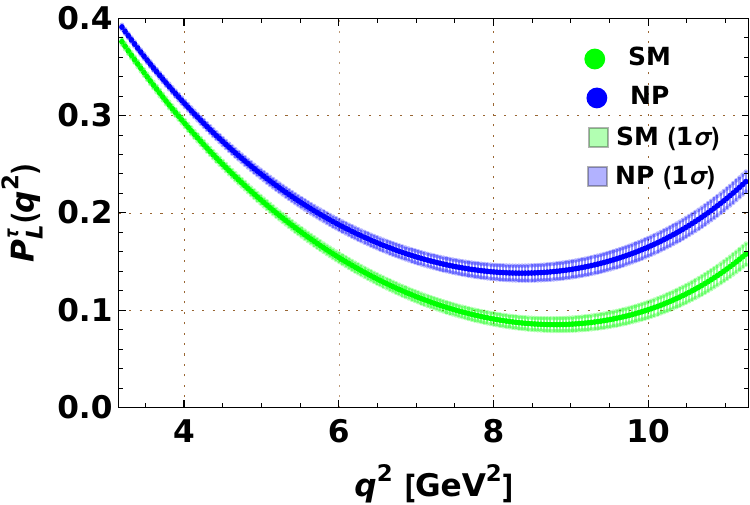}
\includegraphics[height=40mm,width=50mm]{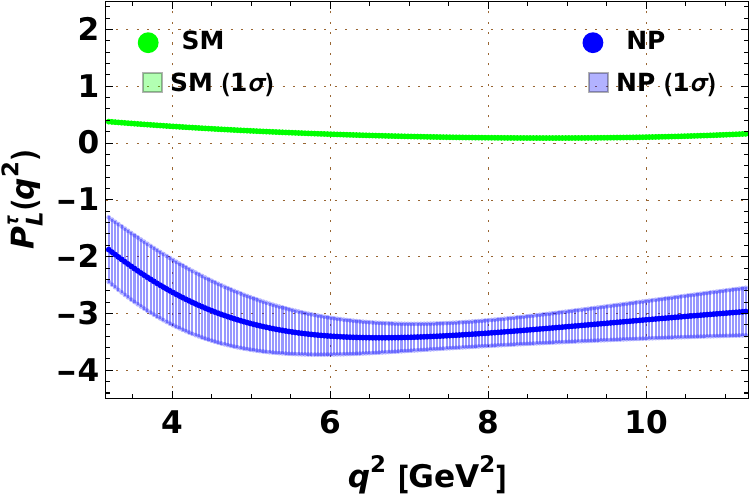}
\caption{$q^2$ dependence of the longitudinal polarisation assymetry  of $\Sigma_b \to \Sigma_c \tau^-\bar{\nu}_\tau$ process. The description of couplings is the same as Fig. \ref{fig::BRCascade}.}
\label{fig::FLSigma}
\end{figure}
 %%%%%%%%%%

The convexity parameter $C_F^\tau$, on the other hand, exhibits a notably large deviation from the SM prediction in the presence of $(C_{lequ}^{(1)}, C_{ledq})$, while a moderate deviation is observed with $(C_{lq}^{(3)}, C_{ledq})$. The details of the $q^2$ behavior is shown in Fig. \ref{fig::CSigma}.

%\ding{43} \textbf{Convexity parameter}:
Much like the convexity parameter $C_F^\tau$, the polarization asymmetry, depicted in Fig. \ref{fig::FLSigma}, shows a significant departure from the SM prediction under the influence of $(C_{lequ}^{(1)}, C_{ledq})$, while a more distinctive deviation is observed with $(C_{lq}^{(3)}, C_{ledq})$.
%\ding{43} \textbf{Polarisation asymmetry}:
%\ding{43} \textbf{Lepton non-universal observable}:
\begin{figure}[h!]
\centering
\includegraphics[height=40mm,width=50mm]{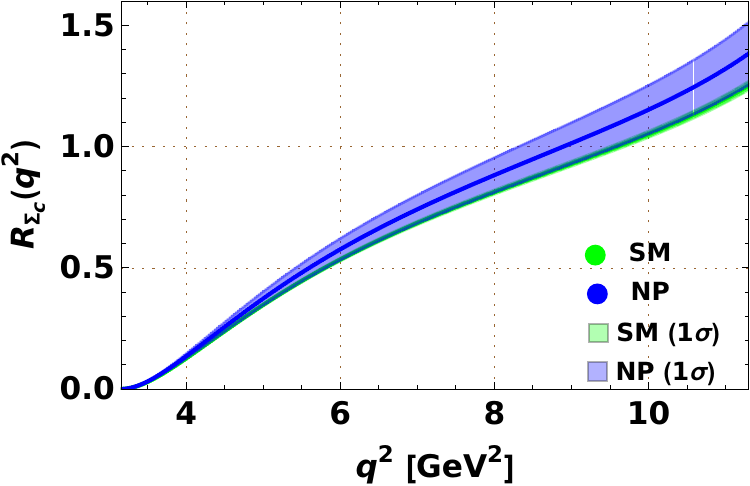}
\includegraphics[height=40mm,width=50mm]{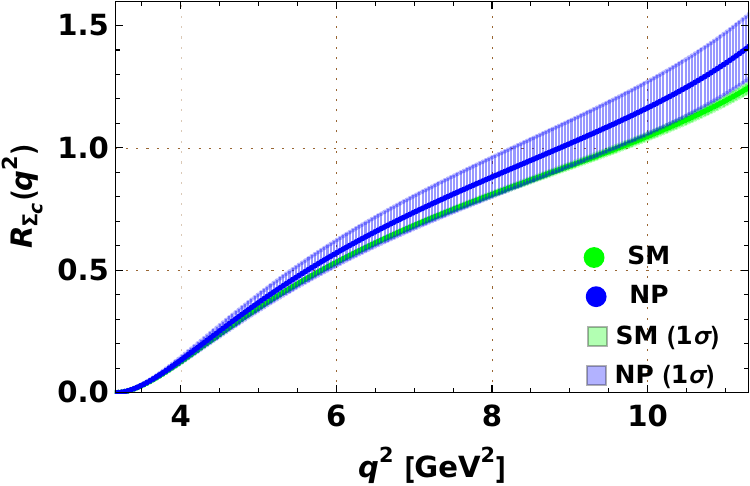}
\includegraphics[height=40mm,width=50mm]{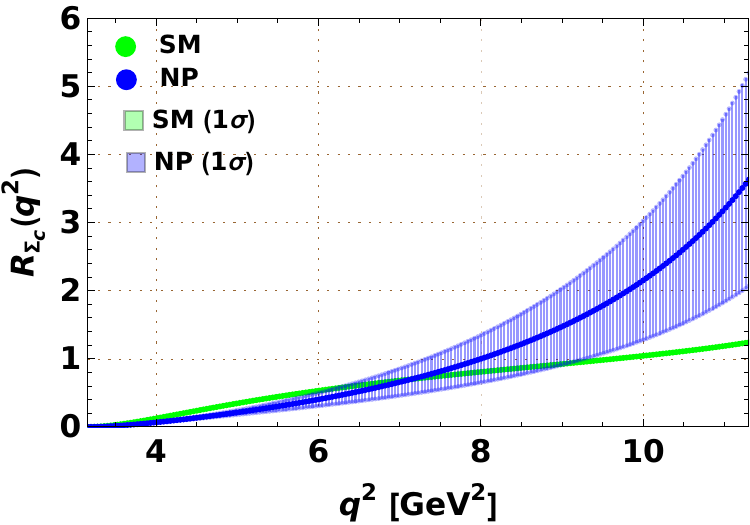}
\caption{$q^2$ behavior of the branching fraction ratio of $\Sigma_b \to \Sigma_c \tau^-\bar{\nu}_\tau$. The description of couplings is same as Fig. \ref{fig::BRCascade}.}
\label{fig::RSigma}
\end{figure}
%%%%%%%%

We further investigate the $q^2$-dependence of the lepton non-universality observable, as shown in Fig.~\ref{fig::RSigma}. A pronounced deviation from the SM prediction is observed at large $q^2$ values, most prominently in the right panel. Conversely, in the low $q^2$ region, the deviation from the SM is minimal. Interestingly, such discrepancies are absent in the other two scenarios: in the left panel, the predictions align closely with the SM, while in the middle panel, a measurable deviation from the SM prediction is observed within the range $q^2 \in [7, 11.3]$~GeV$^2$, suggesting potential contributions from new physics.
%%%%%%%%%%%%%%%%%%%%%%%%%%%%%%
\begin{figure}[h!]
\centering
\includegraphics[height=40mm,width=50mm]{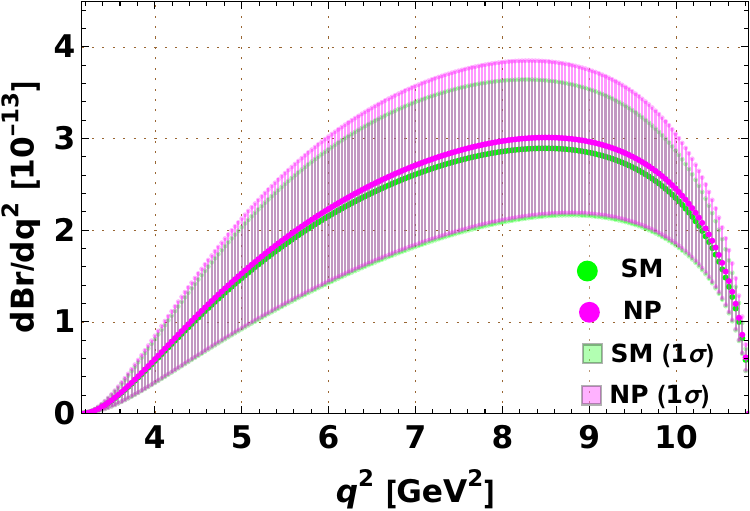}
\includegraphics[height=40mm,width=50mm]{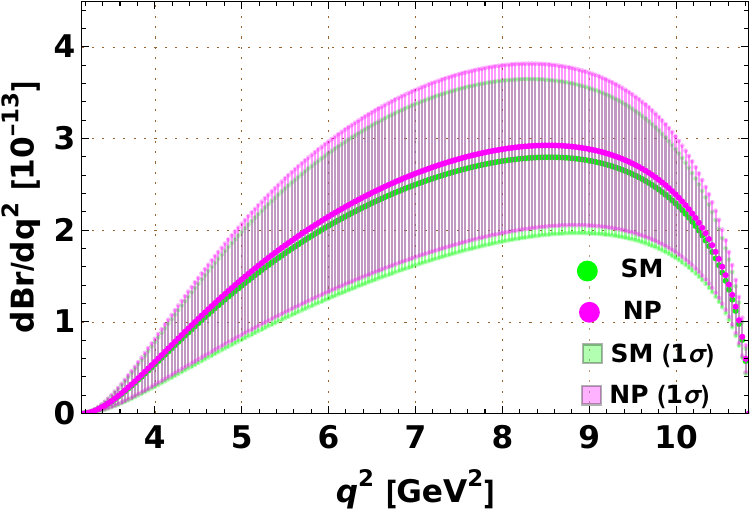}
\includegraphics[height=40mm,width=50mm]{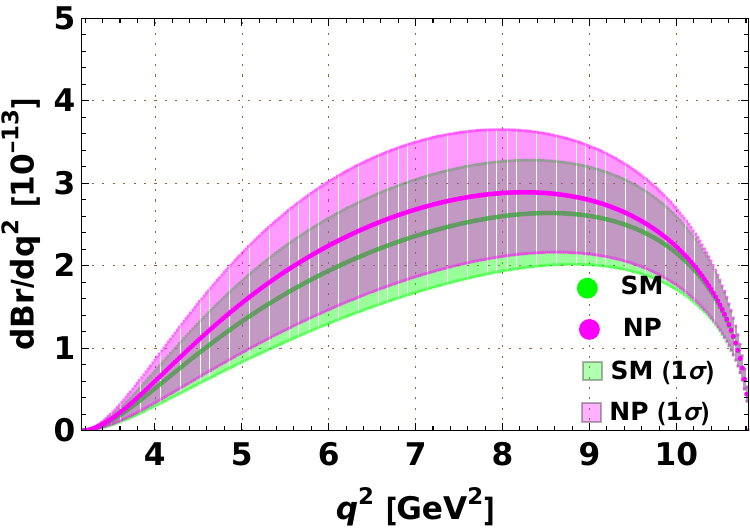}
\quad
\includegraphics[height=40mm,width=50mm]{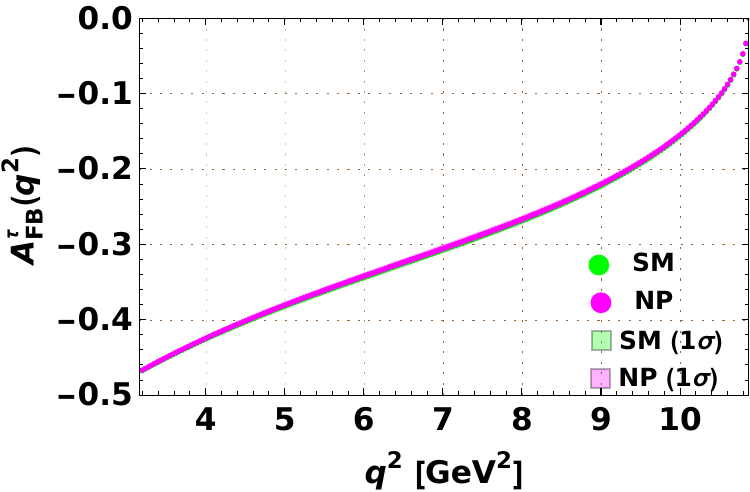}
\includegraphics[height=40mm,width=50mm]{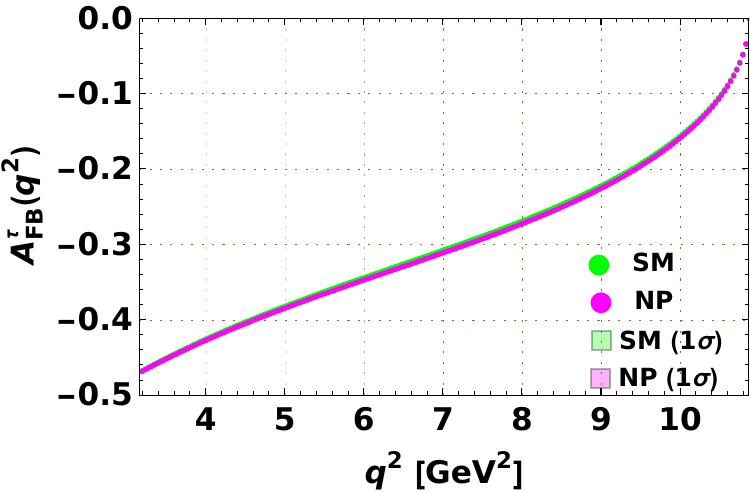}
\quad
\includegraphics[height=40mm,width=50mm]{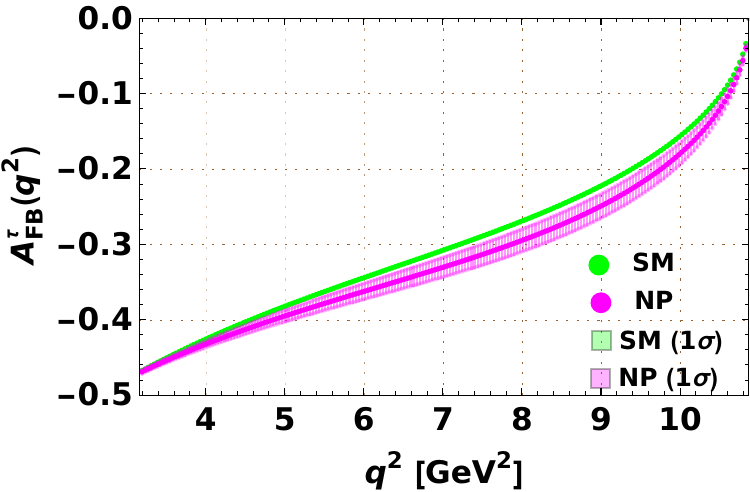}
\caption{The $q^2$ behavior of the differential branching ratio and forward-backward asymmetry of $\Sigma_b \to \Sigma_c^{*} \tau^-\bar{\nu}_\tau$ process. The description of couplings is the same as Fig. \ref{fig::BRCascade}.}
\label{fig::BRSigmastar}
\end{figure}
%%%%%%%%%%%%%%
\subsection{Analysis of $\Sigma_b \to \Sigma_c^{*} \tau^-\bar{\nu}_\tau$ Process}
%%%%%%%%%%%%%
Here, we will discuss the  $\Sigma_b \to \Sigma_c^{*} \tau^-\bar{\nu}_\tau$ decay observables in the presence of NP couplings. Our analysis on the $q^2$ behavior is discussed below.

In the presence of the 2D operators, the branching ratio remains consistent with the SM prediction. On the other hand, the forward-backward asymmetry does not exhibit a zero-crossing point in any of the scenarios. Notably, the couplings $(C_{lequ}^{(1)}, C_{ledq})$ display a distinction both in the SM and within the SMEFT formalism. The rest of the scenarios are almost consistent with the SM prediction. This is presented in Fig. \ref{fig::BRSigmastar}.

\begin{figure}[h!]
\centering
\includegraphics[height=40mm,width=50mm]{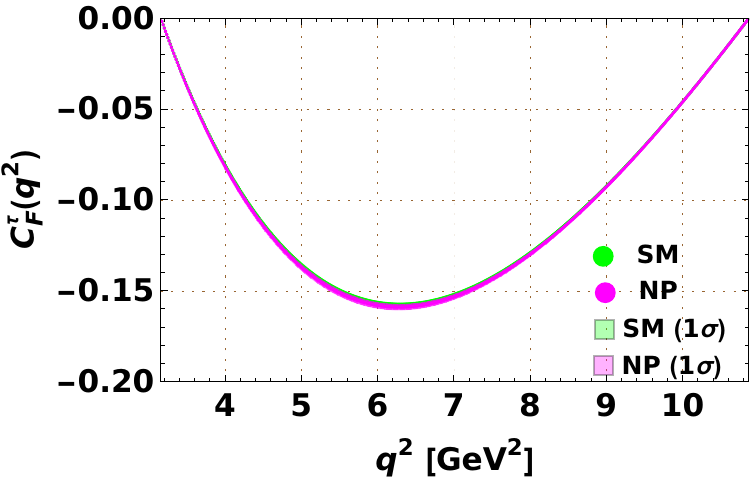}
\quad
\includegraphics[height=40mm,width=50mm]{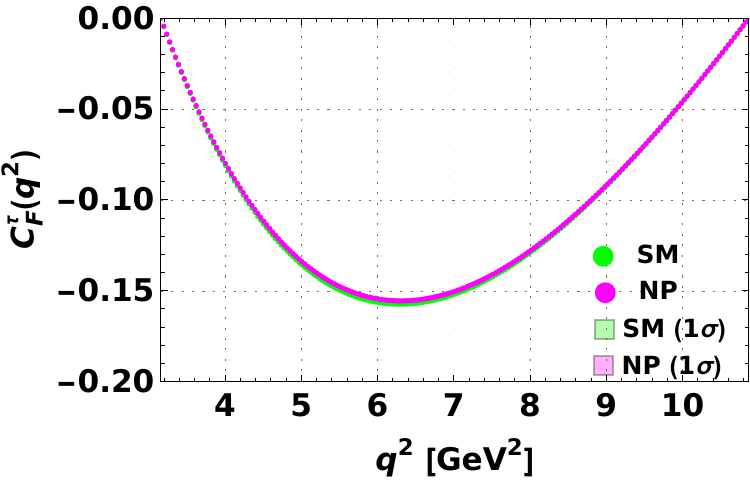}
\includegraphics[height=40mm,width=50mm]{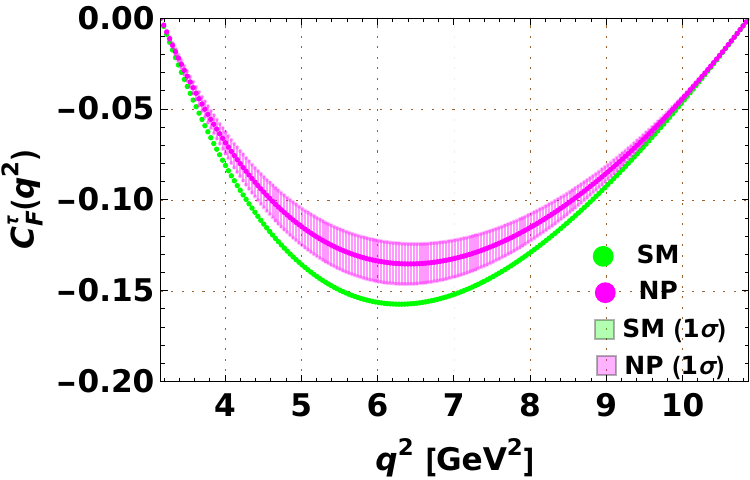}
\caption{Behavior of convexity parameter. The description of couplings is the same as Fig. \ref{fig::BRCascade}.}
\label{fig::CSIgmastar}
\end{figure}
%%%%%%%%%%%%
\begin{figure}[h!]
\centering
\includegraphics[height=40mm,width=50mm]{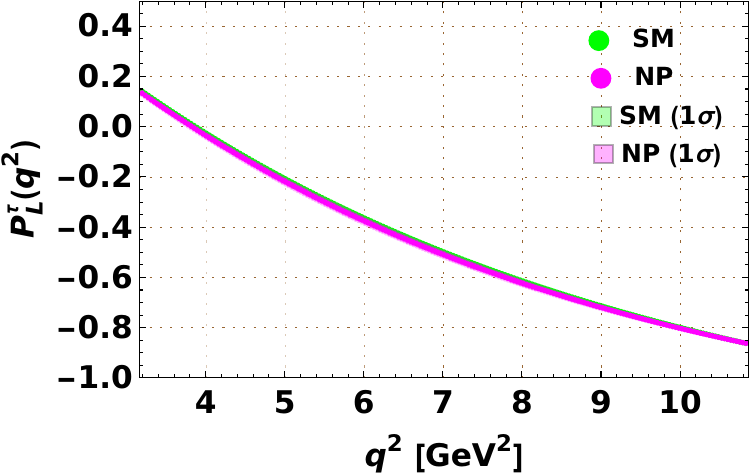}
\quad
\includegraphics[height=40mm,width=50mm]{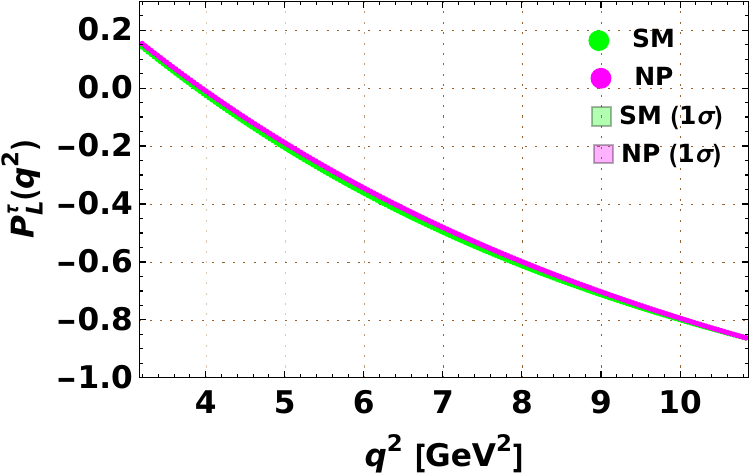}
\includegraphics[height=40mm,width=50mm]{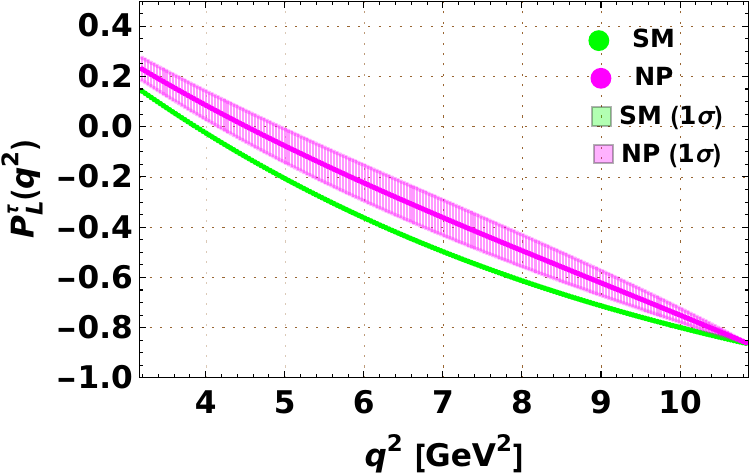}
\caption{Behavior of $P_L^\tau(q^2)$ of $\Sigma_b \to \Sigma_c^{*} \tau^-\bar{\nu}_\tau$ channel. The description of couplings is the same as Fig. \ref{fig::BRCascade}.}
\label{fig::PLSigmastarM}
\end{figure}

%%%%%%%
The convexity parameter $C_F^\tau$ shows a significant deviation from the SM prediction when the couplings $(C_{lequ}^{(1)}, C_{ledq})$ are present, whereas no new physics contribution is seen with the other two scenarios. The detailed $q^2$ behavior is presented in Fig. \ref{fig::CSIgmastar}.

In Fig. \ref{fig::PLSigmastarM}, we observed that the NP sensitivity allows noticeable discrepancies from the SM contribution with the simultaneous presence of $C_{lequ}^{(1)}$ and $C_{ledq}$ couplings.

\begin{figure}[h!]
\centering
\includegraphics[height=40mm,width=50mm]{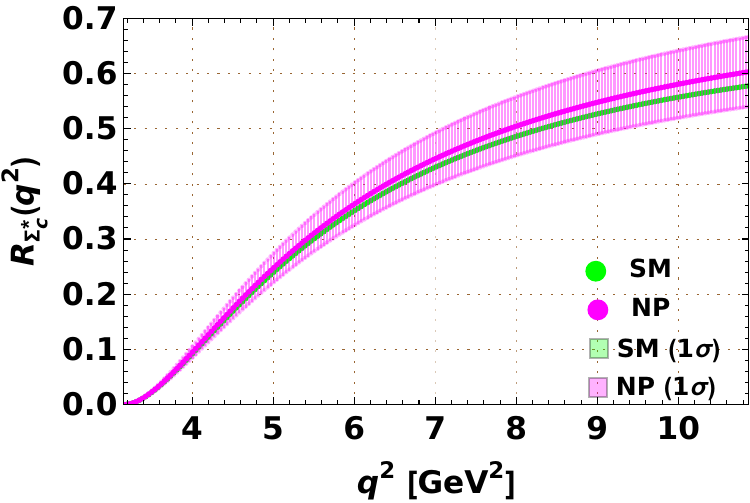}
\quad
\includegraphics[height=40mm,width=50mm]{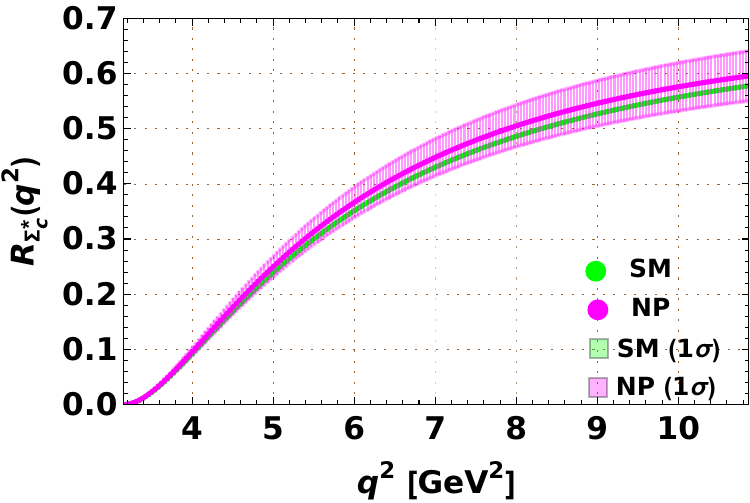}
\includegraphics[height=40mm,width=50mm]{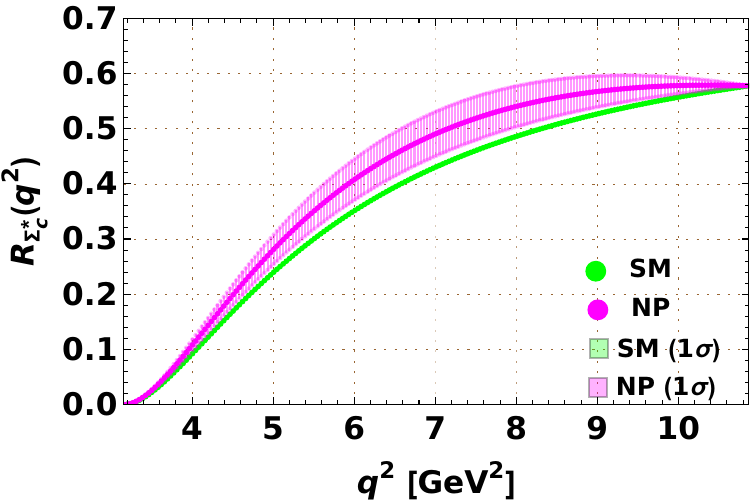}
\caption{The $q^2$ dependent braching fraction ratio of $\Sigma_b \to \Sigma_c^{*} \tau^-\bar{\nu}_\tau$ decay channel. The description of couplings is the same as Fig. \ref{fig::BRCascade}}
\label{fig::RSigmastar}
\end{figure}

In the analysis of the lepton non-universality observable, as illustrated in Fig.~\ref{fig::RSigmastar}, a measurable difference is observed in the presence of the new physics couplings $(C_{lequ}^{(1)}, C_{ledq})$). However, in the presence of other scenarios i.e $(C_{lq}^{(3)}, C_{lequ}^{(1)})$ and $(C_{lq}^{(3)}, C_{ledq})$,  a modest departure is seen in the higher $q^2$ region.

%%%%%%%%%%%%%%%%%%%%%%%%%%%%%%%%%%%
%%%%%%%%%%%%%%%%%%%%%%%%%%%%%%%%%%
\section{Conclusion}
\label{sec:conclu}
%%%%%%%%%%%%%

Motivated by the anomalies observed in charged-current mediated $b \to c \ell \nu$ transitions, we have conducted a comprehensive analysis of the exclusive semileptonic $b$-baryonic decay modes $\Sigma_b \to \Sigma_c^{(*)} \tau^- \bar{\nu}_\tau$ and $\Xi_b \to \Xi_c \tau^- \bar{\nu}_\tau$ within the framework of the SMEFT approach. These decay modes provide a crucial testing ground for probing NP effects.
Using the Weak Effective Theory Lagrangian to describe $b \to c \ell \nu$ transitions, we established explicit correlations between the NP couplings and SMEFT Wilson coefficients. In this context, the SMEFT operators responsible for generating $b \to c \tau \nu_\tau$ transitions also induce LEFT operators relevant to $b \to s \tau^+ \tau^-$ and $b \to s \nu \nu$ processes. This dual impact emphasizes the interconnected nature of these decay channels in the presence of NP contributions. 

For the NP analysis, we constrain the parameter space of the SMEFT couplings, utilizing the experimental data on several key observables, including $R_{D^{(*)}}$, $P_\tau(D^*)$, $F_L(D^*)$, $R_{\Lambda_c}$, $\text{BR}(B \to K^{(*)} \nu \bar{\nu})$, $\text{BR}(B_s \to \tau^+ \tau^-)$, and $\text{BR}(B \to K \tau \tau)$. For our analysis, we assumed the NP couplings to be real. Both 1D and 2D scenarios for the SMEFT couplings were initially considered using the new  Wilson coefficients $C_{lq}^{(3)}$, $C_{lequ}^{(1)}$, and $C_{ledq}$.
However, we found that the contributions of the 1D scenarios were negligible and, therefore, we have not included them in our analysis.

We conducted a detailed examination of the impact of various 2D NP couplings, such as $(C_{lq}^{(3)}, C_{lequ}^{(1)})$, $(C_{lq}^{(3)}, C_{ledq})$, and $(C_{lequ}^{(1)}, C_{ledq})$, on the branching fractions and angular observables, including lepton polarization asymmetry, forward-backward asymmetry, convexity parameter, and the lepton non-universality observable of the $b$-baryonic decay modes $\Sigma_b \to \Sigma_c^{(*)} \tau^- \bar{\nu}_\tau$ and $\Xi_b \to \Xi_c \tau^- \bar{\nu}_\tau$. These results highlight the sensitivity of these processes to NP effects and their potential to complement the insights gained from mesonic decays.

Our results reveal a significant sensitivity to NP effects in $\Xi_b \to \Xi_c \tau \nu$ and $\Sigma_b \to \Sigma_c^{(*)} \tau \bar{\nu}$, with the couplings $(C_{lequ}^{(1)}, C_{ledq})$ yielding more pronounced deviations from the SM predictions than the other couplings. However, the coupling $(C_{lq}^{(3)}, C_{ledq})$ exhibits a moderate deviation in the high $q^2$ region for the convexity parameter, lepton polarization asymmetry, and the lepton non-universality observable of the $\Sigma_b \to \Sigma_c \tau \bar{\nu}$ process.

In summary, this detailed analysis of the $b$-baryonic decay modes $\Sigma_b \to \Sigma_c^{(*)} \tau^- \bar{\nu}_\tau$ and $\Xi_b \to \Xi_c \tau^- \bar{\nu}_\tau$ reveals that the associated observables, both within the Standard Model and in new physics scenarios. 
%Measurements of these observables will offer valuable insights into the impact of new physics on exclusive $b$-baryonic decays mediated by $b \to c \ell \nu$ transitions. 
Despite the experimental challenges associated with these decays, the advancements in detector technology with increasing luminosity could make these channels promising candidates for future measurements.

%%%%%%%%%%%%%%%%%%%%%%%%%%%%%%%%%%%%%%%%%%%%%%%%%%%%%%%%%%%%%%%%%%%%%%%%%%%%%%%%%%%%%%%%%%%%

%

%

%%%%%%%%%%%%%%%%%%%%%%%%%%%%%%%%

%%%%

%

%%%%%%

%%%%%%%%%%%%%%%%%%%%%%%%%%%%%%%%%%%%%%%%

%%%%%%%%%%%%%%%%%%%%%%%%%%%%%%%%%%%%%%%%%%%%%%%%%%%%%%%%%%%%%%%%

%%%%%%%%%%%%%%%%%%%%%%%%%%%%%%%%%%%%%%%%%%%%%%%%%%%%%%%%%%%%%%%%%%%%%%%%%%%%%%%%%%%%%%%%%%%%%%
\section{Acknowledgement}
DP would like to acknowledge the support of the Prime Minister's Research Fellowship, Government of India. MKM acknowledges the financial support from IoE PDRF, University of Hyderabad. RM would like to acknowledge the University of Hyderabad IoE project grant no. RC1-20-012.\\

{\bf Data Availability Statement} This manuscript has no associated data.

%{\bf Declarations}

%{\bf Conflict of interest} The authors declare no conflict of interest.
%===================================================
%%%%%%%%%%%%%%%%%%%%%%%%%%%%%%%%%

%%%%%%%%%%%%%%%%%%%%%%%

%%%%%%%%%%%%%%%%%%%%%%%%%%%%%

%%%%%%%%%%%%%%%%%%%%%%%%%%%

%%%%%%%%%%%%%%%%%%%%%%%%%%%%%%%%%%%%%%%%%%%%%%%%%%%%%%%%%

%%%%%%%%%%%%%%%%%%%%%%%%
\bibliographystyle{ieeetr}
\bibliography{RDM}

%%%%%%%%%%%%%%%%%%%%%%%%%%%%
\end{document}